%% file: main.tex
\documentclass[final,5p,twocolumn,10pt]{elsarticle} 

\include{preamble}

\graphicspath{{figures/}}
\journal{Arxiv}

\begin{document}

\begin{frontmatter}



\title{PATO: Producibility-Aware Topology Optimization using Deep Learning for Metal Additive Manufacturing}

\author[GER]{Naresh S. Iyer\corref{contrib}}
\ead{iyerna@ge.com}
\author[PARC]{Amir M. Mirzendehdel\corref{contrib}}
\ead{amirzend@parc.com}
\author[GER]{Sathyanarayanan Raghavan\corref{contrib}}
\author[GER]{Yang Jiao}
\author[PARC]{Erva Ulu}
\author[PARC]{Morad Behandish}
\author[PARC]{Saigopal Nelaturi}
\author[GER]{Dean M. Robinson}

\cortext[contrib]{Authors contributed equally}
\address[GER]{GE Research (GER), 1 Research Circle, Niskayuna, NY, United States}
\address[PARC]{Palo Alto Research Center (PARC), 3333 Coyote Hill Rd., Palo Alto, CA, United States}

\input{Abstract}

\end{frontmatter}


\input{Introduction}
\input{LiteratureReview}
\input{Problem}

\input{ProposedMethod}

\input{Results}
\input{Conclusion}
\input{Acknowledgements}
\bibliographystyle{elsarticle-num} 
\bibliography{cas-refs.bib}

\end{document}

%% file: preamble.tex
\usepackage{hyperref}
\usepackage{enumitem}

\usepackage{amsmath}
\usepackage{amssymb}
\usepackage{stmaryrd}
\usepackage{amsthm}
\usepackage{amsfonts}
\usepackage{dsfont}
\usepackage{graphicx}
\usepackage{upgreek}
\usepackage{bm}
\usepackage{float}
\usepackage{tikz-cd}
\usepackage{relsize}

\usepackage{mathtools}
\usepackage{algorithm,algpseudocode} 

\usepackage{todonotes}

\usepackage{subcaption}

\usepackage{tabu}
\usepackage{dblfloatfix} 

\usepackage{graphicx}
\usepackage{graphbox}

\usepackage{color}

\biboptions{numbers,sort&compress}

\newcommand{\bu}{{\mathbf{u}}}

\newcommand{\bK}{{\mathbf{K}}}
\newcommand{\bff}{{\mathbf{f}}}

\newcommand{\MSSImax}{\zeta}

\newcommand{\targ}{\text{target}}

\theoremstyle{definition}

\newcommand{\com}[1]{} 

%% file: Abstract.tex
\begin{abstract}
In this paper, we propose PATO---a producibility-aware topology optimization (TO) framework to help efficiently explore the design space of components fabricated using metal additive manufacturing (AM), while ensuring manufacturability with respect to cracking. Specifically, parts fabricated through Laser Powder Bed Fusion (LPBF) are prone to defects such as warpage or cracking due to high residual stress values generated from the steep thermal gradients produced during the build process. 
Maturing the design for such parts and planning their fabrication can span months to years, often involving multiple handoffs between design and manufacturing engineers. PATO is based on the {\it a priori} discovery of crack-free designs, so that the optimized part can be built defect-free at the outset. To ensure that the design is crack free during optimization, producibility is explicitly encoded within the standard formulation of TO, using a crack index. Multiple crack indices are explored and using experimental validation, maximum shear strain index (MSSI) is shown to be an accurate crack index. Simulating the build process, in order to estimate MSSI, is a coupled, multi-physics, time-complex computation and incorporating it in the TO loop can be computationally prohibitive. We leverage the current advances in deep convolutional neural networks (DCNN) and present a high-fidelity surrogate model based on an Attention-based U-Net architecture to predict the MSSI values as a spatially varying field over the part's domain.
Further, we employ automatic differentiation to directly compute the gradient of maximum MSSI with respect to the input design variables and augment it with the performance-based sensitivity field to optimize the design while considering the trade-off between weight, manufacturability, and functionality. We demonstrate the effectiveness of the proposed method through benchmark studies in 3D as well as experimental validation.  
\end{abstract}


%% file: Introduction.tex
\section{Introduction}
\begin{figure*} [ht!]
	\centering
	\includegraphics[width=\linewidth]{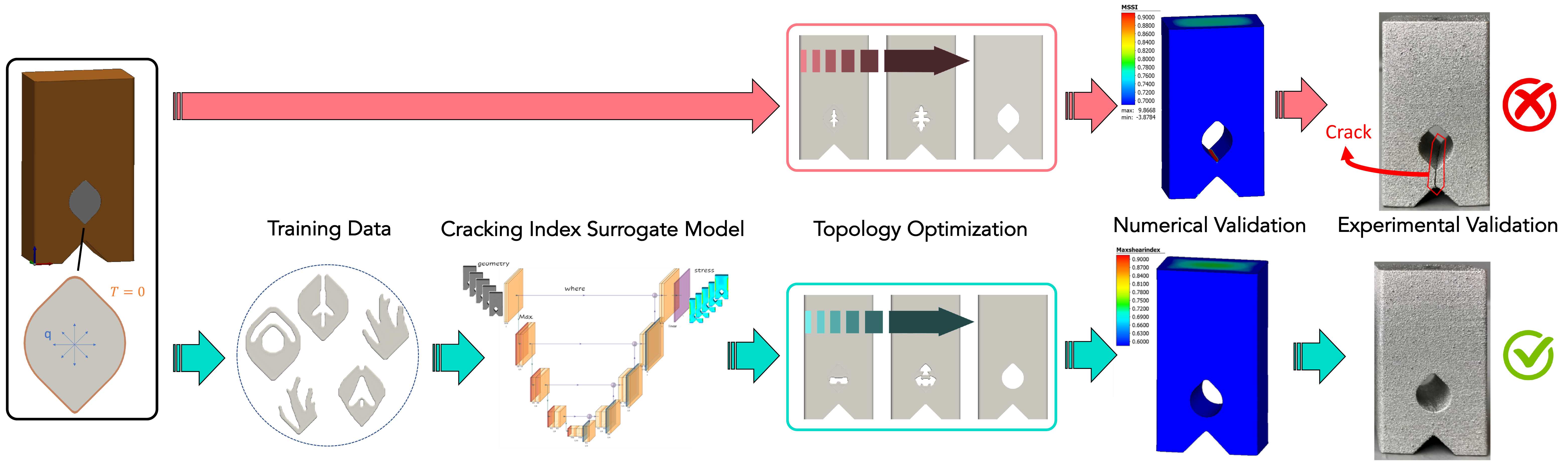}
	\caption{Our producibility-aware topology optimization (PATO) framework enables an approach to incorporate producibility of additively manufactured parts directly into the design process using a deep neural network predictor of producibility. For instance, given a heat conduction problem (left), a typical TO approach generates `no-go' designs with regions of high residual stress that crack after manufacturing (top). On the other hand, PATO leverages an attention-based neural network surrogate model trained on a diverse set of optimized designs to augment the performance sensitivity field with a producibility sensitivity field (here MSSI) obtained through automatic differentiation to generate `go' designs that will remain crack-free after post-processing (bottom).}
	\label{fig_CoverFig}
\end{figure*}

In this paper, we explore design synthesis of parts fabricated by metal additive manufacturing (AM) while considering `producibility' constraints. For the purposes of this paper, producibility is defined by the ability to fabricate the part such that there are no cracks produced due to steep thermal gradients encountered during LPBF-AM. 
Solving this problem is critical to design gas turbine components, especially those that are equipped with channels designed for hot gas flow, and fabricated using LPBF-AM of Ni-based superalloys~\cite{debroy_additive_2018}.
In LPBF-AM, a focused laser beam is used to selectively melt powder in a layer-by-layer fashion~\cite{Frazier2014MetalAM} to construct a geometry of interest. Each layer can be a few tens of microns thick and after melting, each layer is cooled in order to solidify it before sequentially depositing the next layer on top. These rapid heating and cooling cycles can lead to steep thermal gradients in the part that can result in generation of high residual stresses. These residual stresses can manifest in part deformation or if sufficiently high, result in cracking at multiple locations of the part~\cite{Khairallah2016LaserPF}. When a part cracks at one or more locations after manufacturing, it becomes necessary to refine the part design and to repeat the manufacturing process with the new design, leading to multiple iterations from design to full scale manufacture of the part. Consequently, the overall design maturation time for a part can span from months to years, involving multiple hand-offs between design and manufacturing engineers. Additionally, as metal AM evolves from manufacturing part prototypes to large scale high volume industrial parts, throughput becomes a critical factor for the adoption of AM technology. Addressing this inefficiency and cost, due to cracking after manufacturing, requires an understanding and modeling of the phenomena that lead to the initiation of cracks. We demonstrate how an understanding of the cracking mechanism as it relates to the melting process in AM can be used to formulate a crack index by leveraging high-fidelity multiphysics simulators of the melting process. We show how an attention-based, 3D deep learning neural network architecture~\cite{Iyer2021AttentionBased3N} can be used to model the crack index accurately and act as a reliable surrogate for the time-complex, high-fidelity simulator. This enables instantaneous estimation of the crack index, and permits an approach for the crack index to be directly incorporated into algorithms that optimize the design of components producible using LPBF-AM.

Topology optimization (TO) is a powerful automated design paradigm that identifies optimized realizations of a conceptual design problem defined in terms of loading conditions, and performance and manufacturing objectives and constraints. It enables efficient exploration of large design spaces, and can harness the true potential and innovation of AM. On the other hand, the geometric complexity of designs generated by TO can cause defects in the manufactured part, which can result in costly trial-and-error cycles. Generating producible (as defined earlier) optimized designs requires explicitly encoding producibility within TO and computing the gradients. We employ automatic differentiation to compute the gradients, a.k.a. the sensitivity field, of the producibility criterion and augment it with the performance sensitivity fields. Subsequently, the augmented sensitivity field is used to optimized the design with respect to both performance and producibility. We demonstrate that closing the loop from manufacturing to design in this manner allows for TO to successfully stay away from crack-prone designs as illustrated in Fig.~\ref{fig_CoverFig}.

\subsection{Contributions \& Outline}

The key contributions of this paper are:

  \begin{itemize}
    \item A producibility-aware topology optimization (PATO) approach to constrain the solution of multi-physics TO and guide it away from non-producible designs;
    \item The application of automatic differentiation to generate sensitivity fields that can be integrated into the standard TO formulation to effectively bias the optimization away from crack-prone design candidates.
    \item The formulation of a specific producibility criterion; namely, propensity to cracks in additively manufactured parts (via LPBF-AM) in terms of a maximum shear strain index (MSSI), whose correlation with cracking is established by experimental testing.
    \item The volumetric regression of the producibility criteria as a function of part geometry via an attention-based, 3D convolutional neural network (CNN) surrogate model, including training and validation.
    %
    %
    \item An intelligent workflow for the selection of the maximally diverse set of design candidates, from those generated by the design generator for training the surrogate; and
\end{itemize}

To demonstrate the effectiveness of our PATO framework, we choose a problem featuring the design of an air cooling channel on a test coupon, representative of the channels encountered in turbomachinery parts. We show that PATO is able to successfully converge towards the discovery of crack-free designs, while a TO algorithm unconstrained by producibility applied to the same problem is unable to accomplish the same outcome.

We identify MSSI as an appropriate crack-index based on an understanding the physics related to the crack initiation process, and use a calibrated inherent strain-based high-fidelity modeling tool of the nonlinear and transient additive melting process to evaluate for train/test data generation for the surrogate model. Noting that the MSSI can be predicted as a function of the part geometry, we use a design generator to construct training data candidates using TO under varying multi-physics constraints. Most of the related traditional efforts only focus on constraints specific to structural integrity, which can render many candidates infeasible in the context of turbomachinery parts due to the criticality of the constraints imposed by other physics-based (e.g., thermal, flow, and pressure) effects. Our design generator further incorporates constraints that ensure that the generated geometries are free from self-support issues during additive fabrication, thereby ensuring that they can be feasibly printed by the additive process.
The selection workflow shortlists the training data to a maximally diverse set of candidates, ensuring a good trade-off between the computational complexity of evaluating the training data with the generalization capability of the surrogate model in the design space.

To efficiently predict cracking for each candidate design generated within the TO loop, we train an attention-based CNN surrogate model \cite{Guo_SAUnet}. We show how neural attention mechanisms help overcome the challenge of detecting extremely sparse regions of crack propensity reliably, accurately, and time-efficiently. In addition to computational efficiency, the CNN architecture facilitates deriving 3D sensitivity fields via automatic differentiation, regardless of the inherent complexity of the original physics-based model used for training data generation.

%% file: LiteratureReview.tex
\section{Literature Review} \label{sec_litReview}
In this section, we will briefly review recent advances in TO for manufacturing, failure criteria to predict cracking due to residual stresses, and deep learning for AM.  

\subsection{Topology Optimization for Manufacturing}
TO~\cite{Bendsoe2009topology,mirzendehdel2017hands} is often used at the preliminary stages of design to provide insight on viable geometric features and material distribution~\cite{mirzendehdel2015pareto} of a product. An unaddressed challenge in the widespread adoption of TO in the final stages of the design cycle is non-manufacturability of the organic and geometrically complex TO shapes. To bridge the gap between design and manufacturing, various manufacturing constraints have been incorporated within TO formulation~\cite{sigmund2009manufacturing,Liu2016survey,Vatanabe2016topology,Zhou2002progress}. Considering some of the traditional manufacturing technologies, a few methods have been proposed that extend the TO formulation to ensure manufacturability by casting~\cite{harzheim2006review,Li2018topology,Wang2017structural,Guest2012casting}, laser-cut~\cite{Vatanabe2016topology,mirzendehdel2017hands}, or milling~\cite{Guest2012casting,mirzendehdel2019exploring,mirzendehdel2020topology,Zuo2006manufacturing,Langelaar2019topology,liu20153d}. On the other hand, AM enables engineers to fabricate highly detailed and complex parts and offers a great synergy with TO. However, different AM processes require distinct design guidelines to be followed for a successful and cost-effective build. Therefore, design for AM (DfAM) and specifically TO for AM is a topic of great interest, where strategies have been proposed for TO with respect to feature size~\cite{Zhou2015minimum}, sacrificial support structures~\cite{gaynor2016topology,Langelaar2016topology,mirzendehdel2016support}, material uncertainty~\cite{chen2010level,lazarov2012topology}, and process-induced anisotropy~\cite{Mirzendehdel2018strength}. In this paper, we focus on TO for LPBF-AM, where a laser beam melts and sinters metal powder in a layer-by-layer fashion to fabricate the part. Such additive processes are well-recognized as one of the most important AM technologies~\cite{debroy_additive_2018} in that they feature minimal part surface roughness, high dimensional accuracy as well as great geometric freedom and versatility. 

On the other hand, the complicated physical processes involved in LPBF-AM pose a major challenge in the efforts to combine the LPBF-enabled fabrication freedom with the TO-informed design optimality. 

In fact, the highly localized and transient input of tremendous energy throughout a LPBF-AM process subjects every region of the part to multiple phase changes in rapid heating and cooling cycles. This results in high thermal gradient and subsequently residual stresses that can give rise to detrimental cracking of the part. In addition, part geometry has been identified as an important factor correlated to the tendency of cracking in a LPBF process~\cite{ghouse_vacuum_2021,lee_correlations_2020}. This indicates the necessity to incorporate a constraint for crack-free AM fabrication into TO so that the AM manufacturability of the resulting optimal design can be ensured. The producibility constraint can potentially be implemented through assessing cracking risk of the designs by means of high-fidelity printing process simulation. However, this approach requires coupled multi-physics, time-dependent, and nonlinear analysis at high-resolution; using it in an optimization loop is currently computationally prohibitive. Further, deriving analytical expressions for the gradients is extremely challenging. Recently, a number of paradigms have been developed to address this issue through acceleration of the TO engine~\cite{kallioras2020accelerated}; another approach directly executes TO by leveraging an machine learning (ML) framework~\cite{chandrasekhar2021tounn,chandrasekhar2021multi,chandrasekhar2021length,chi2021universal}. We leverage recent advances in ML to efficiently predict cracking within the TO using a surrogate model as well as computing gradients using automatic differentiation. 

\subsection{Metrics for predicting cracks}
Metal alloys typically fail due to nucleation, growth and coalescence of microscopic defects/voids. Based on experimental observations, analytical studies and numerical modeling, several failure parameters have been proposed in the literature. Gurson~\cite{gurson1977continuum, gurson1975plastic} published seminal work in developing a model to predict failure of ductile materials. The model involved more than ten parameters that needed to be calibrated independently. Needleman and Tvergaard~\cite{tvergaard1981influence,tvergaard1982localization} completed the model and proposed methods to calibrate the parameters. With the advent of finite-element modeling, it became easier to estimate complex parameters such as stress triaxiality that characterizes relative degree of hydrostatic stress in a given stress state. Therefore, researchers~\cite{bao2004fracture, bai2008new,lou2012new} started considering various stages of crack growth such as void nucleation, growth, coalescence independently and described each phenomenon as a function of equivalent plastic strain, stress triaxiality, normalized maximal shear stress respectively.

In this work, we explored applicability of three parameters, which are described by Bao et. al.~\cite{bao2004comparative} - strain failure index (SFI), maximum shear strain index (MSSI) and total strain energy density index (TSI), as criteria to predict crack likelihood.

Through experimental coupon-builds and calibration, we concluded that the MSSI serves as a good index to capture likelihood of crack with suitable precision and specificity.

\subsection{Deep Learning for prediction of dense 3D fields}
The application of 3D segmentation techniques using convolutional neural networks (CNNs), while prevalent in medical imaging domain~\cite{iek20163DUL,Milletari2016VNetFC,Lee2017SuperhumanAO,Yu2017VolumetricCW,Zhou2018PerformanceEO,Ghavami2019AutomaticSO}, is fairly new in the domain of additive manufacturing. Many of the workflows implement a 2D based inference followed by postprocessing to stitch the outcomes volumetrically. Qi et. al.~\cite{Qi2019ApplyingNM} provided a broad survey into the application of deep learning to AM. The work described in ~\cite{Scime2018AMC} is analogous to ours, where a 3D U-Net is applied for segmentation of 3D printed volumes to facilitate automated identification of defects in the part. Unlike our paper, this work targets the standard segmentation task, formulated as classification and does not need to target the voxel-level spatial resolution for regression that is critical for the problem targeted in our paper. Some other related efforts in the space of AM include~\cite{Scime2018AMC,Zhou2018PerformanceEO,korneev2020fabricated} that make use of 2D inference of defects during AM, by analysis of 2D camera images during part printing. Khadilkar et. al.~\cite{Khadilkar2019DeepLS} considered stress prediction for AM parts; while it employed high-fidelity physics-based simulation, and a deep learning based model as a surrogate, to estimate stress for varying geometries, it largely focused on modeling 2D separation stress at the interface that occurs in bottom-up stereolithography printing. Liang et. al.~\cite{Liang2018ADL} used finite element model and deep learning to estimate surface Von Mises stress distribution on aorta walls; however, it also abstracts the estimation problem into a 2D modeling problem by unrolling the aorta wall into a 2D surface, by applying a shape abstraction model. Nie et. al.~\cite{Nie2020StressFP} presented the  application of deep neural networks for predicting the 2D stress fields on cantilevered structures. To the best of our knowledge, our work described in~\cite{Iyer2021AttentionBased3N} is the first body of work to look at the application of attention-based 3D architectures for a full end to end volumetric regression of 3D stress fields in AM. This paper will extend the outcomes of~\cite{Iyer2021AttentionBased3N} for the accurate estimation of  crack index from 3D geometry of a design candidate.

%% file: Problem.tex
\section{Design Problem} \label{sec_problem}

\begin{figure} [ht!]
	\centering
	\includegraphics[width=0.75\linewidth]{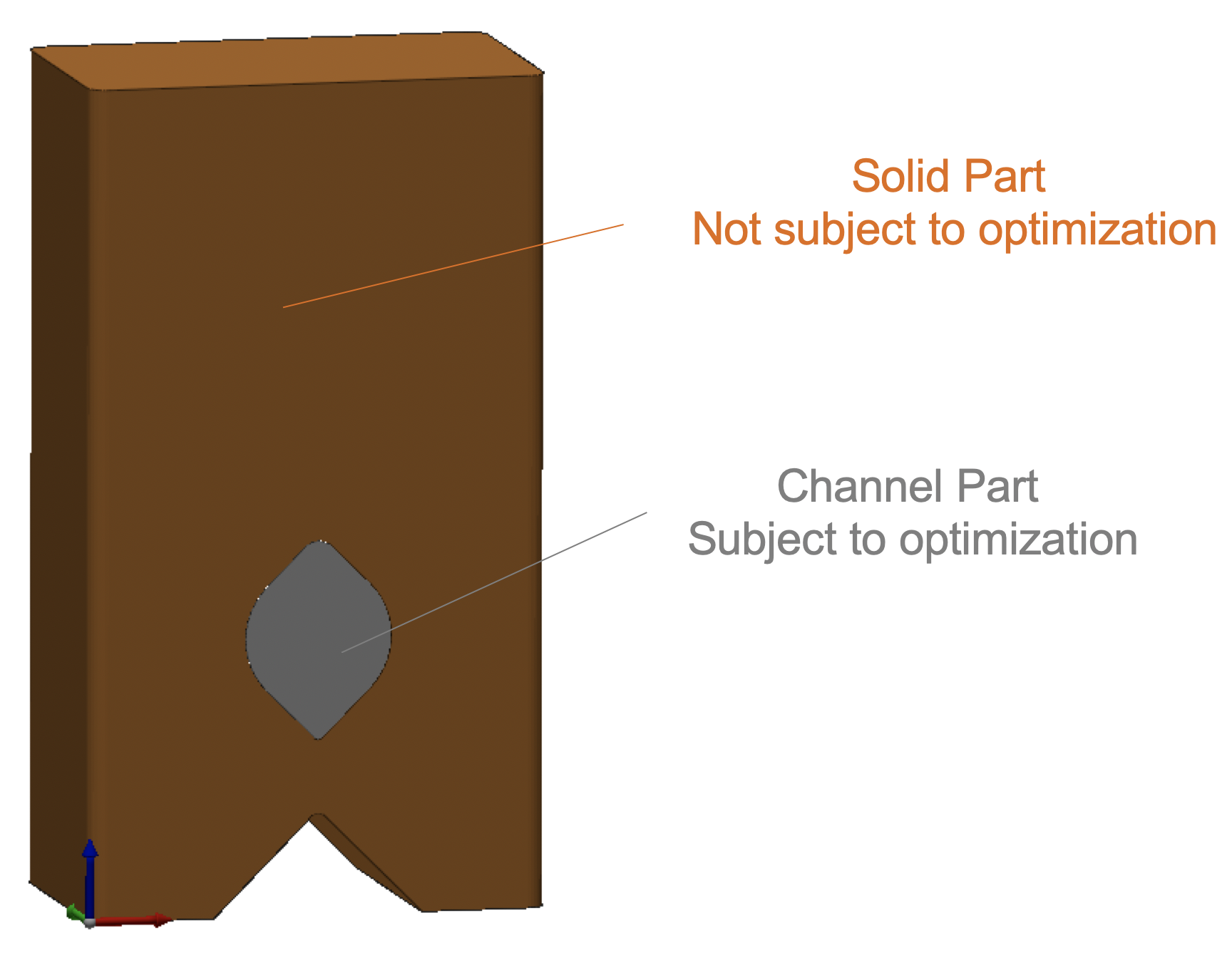}
	\caption{No-Go coupon design domain.}
	\label{fig_nogoDesignProblem}
\end{figure}

The design problem targeted in this paper is the same as the one described in ~\cite{Iyer2021AttentionBased3N}. It deals with the design of a channel, which can be seen either as a cooling channel for hot gas flow or a mechanism to help reduce weight in turbomachinery parts. To emulate this design problem, a coupon as shown in Fig.~\ref{fig_nogoDesignProblem} was used to help formulate the design space. 

In addition to the cooling channel, the coupon includes a triangular notch at the bottom to emulate presence of another geometric artifact in the vicinity of the channel, that might exist in the actual part. In our experiment, the primary parameter is the shape and volume of the channel. Additional design and manufacturing constraints include:
\begin{itemize}[noitemsep,topsep=0pt,parsep=0pt,partopsep=0pt]
\item Minimum feature size of 0.5 mm
\item Symmetry in X direction
\item Extrusion in Y direction to ensure 1) air flow and 2) removal of excess powder
\item Self-supporting in Z direction with $45^\circ$ overhang angle
\end{itemize}
Our experiments show that many variants of this configuration consistently lead to cracks after fabrication. The goal is to understand the relationship between the channel shape, volume, and the crack index distribution, so that designs that lead to high crack index values can be avoided at the outset during the design phase. While approaches like transfer learning~\cite{RoyAM3} can facilitate extending the surrogate model generated for a given feature to other design problems that are only incrementally different, or to varying feature-sizes, for a drastically different design feature or geometry, a new surrogate might need to be created emulating the workflow presented in this paper. In that sense, the vision is for there to be a library of surrogates each of which would apply to a distinctly different feature and help optimize it for crack-free manufacturing, within a class of design problems.

%% file: ProposedMethod.tex
\section{Proposed Method} \label{sec_proposedMethod}

In this section, we will describe the proposed method by explaining the process modeling and inherent strain calibration (Section~\ref{subsec_processModel}), failure prediction through various cracking indices (Section~\ref{subsec_mssi}), training data generation using TO with different objectives and constraints (Section~\ref{sec_trainingdata}), selection of maximally diverse design samples (Section~\ref{sec_trainingSelect}), evaluation of the training data numerically (Section~\ref{sec_trainingEval}), attention-based deep learning surrogate model in 3D for cracking index (Sections~\ref{sec_surrogateOverview} and~\ref{sec_surrogateOutcome}), and the PATO formulation and sensitivity analysis by leveraging automatic differentiation (Section~\ref{sec_TOformulate}).

\subsection{Additive Manufacturing - Process Modeling}\label{subsec_processModel}
\begin{figure} [h!]
	\centering
	\includegraphics[width=\linewidth]{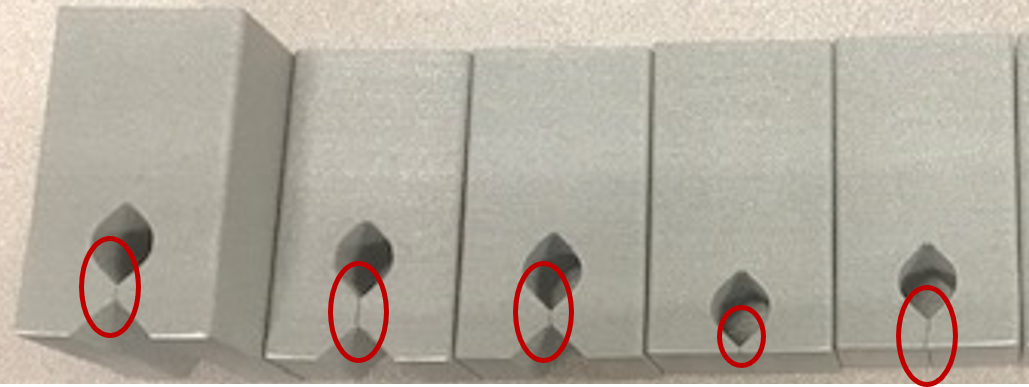}
	\caption{Multiple trials of the baseline coupons that crack repeatedly at the base of the notch (black ellipses show cracks).}
	\label{fig_nogos}
\end{figure}

The  baseline coupon is designed through prior knowledge and several experimental trials wherein the coupon cracks repeatedly at the base of the notch, as shown in Fig. \ref{fig_nogos}. Such a coupon mimics a crack observed in typical 3D printed hot gas path component in jet engines. Addressing the cracking problem requires understanding and accurate modeling of the phenomena that results in the initiation of the cracks. Since cracks are a mechanism to relieve stresses, residual stresses at the end of build process is studied in this work. 

Several commercially available high-fidelity and physics-based software tools can simulate the nonlinear, transient, and multiphysics additive build process. Although such multiphysics simulations are attractive to study the process and capture the physics accurately, they tend to be resource intensive and take a long time to solve. For example, the process simulation of a 7-inch tall part can take more than a week on a 24 core high performance processor with 256 GB of memory. Multiple approaches attempt to address defect control as a process control problem, wherein undesirable change to one or more of the process parameters is inferred using a mix of physics-based and sensor-based models, to then adjust the process parameters to towards the desirable regime of values~\cite{BlomAM, RoyAM1, RoyAM2, IyerAM}. The goal of this study is to modify the design without modifying the process parameters because the feasible process window to successfully print parts using Ni superalloys is narrow. Therefore, the inherent strain methodology is used to simulate the build process. Such a technique is several orders of magnitude faster in-terms of compute time than the multiphysics simulation. However, the inherent strains need to be predetermined. 

\begin{figure} [ht!]
	\centering
	\includegraphics[width=0.9\linewidth]{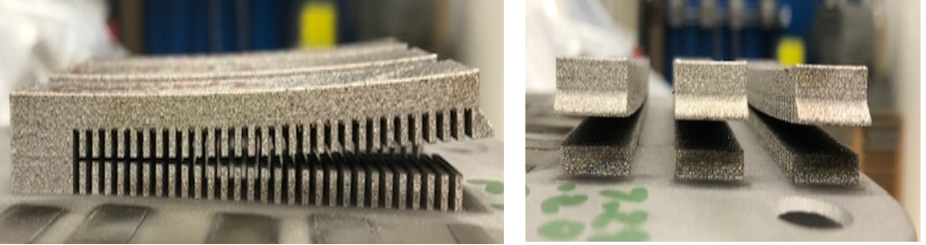}
	\caption{Inherent strain calibration using cantilever coupons.}
	\label{fig_calibrate}
\end{figure}

There are several methods to calibrate the inherent strains in order to predict the stresses accurately. In this study inherent strain calibration based on experiments is pursued. Three cantilever coupons are printed and released using electric discharge machining (EDM) process. The resulting out-of-plane displacement (parallel to build direction) is measured at several locations (see Fig. ~\ref{fig_calibrate}). The calibration process involves simulating the build process and varying the inherent strains until the predicted distortion matches the measured distortion at various locations. The calibrated inherent strain values were found to be $\epsilon_{xx}=\epsilon_{yy}=-0.010295$ and $\epsilon_{zz}=-0.03$. If the process parameters or material are changed, then the calibration process will need to be repeated. 
Apart from computational efficiency, another advantage of the inherent strain based methodology over the coupled transient thermal-structural model is that the characterization of temperature-dependent material properties is not required.
Only the room temperature mechanical properties such as modulus, Poisson's ratio and flow stress versus plastic strain is sufficient to fully define the material. 
The inherent strain based process modeling approach predicts residual stress gradients fairly accurately which are the primary drivers of the crack. Therefore, the predicted residual stresses can be used to synthesize an appropriate 3D crack index that captures the likelihood of crack across the entire part. 

\subsection{Crack Index}\label{subsec_mssi}

\begin{figure} [ht!]
	\centering
	\includegraphics[width=\linewidth]{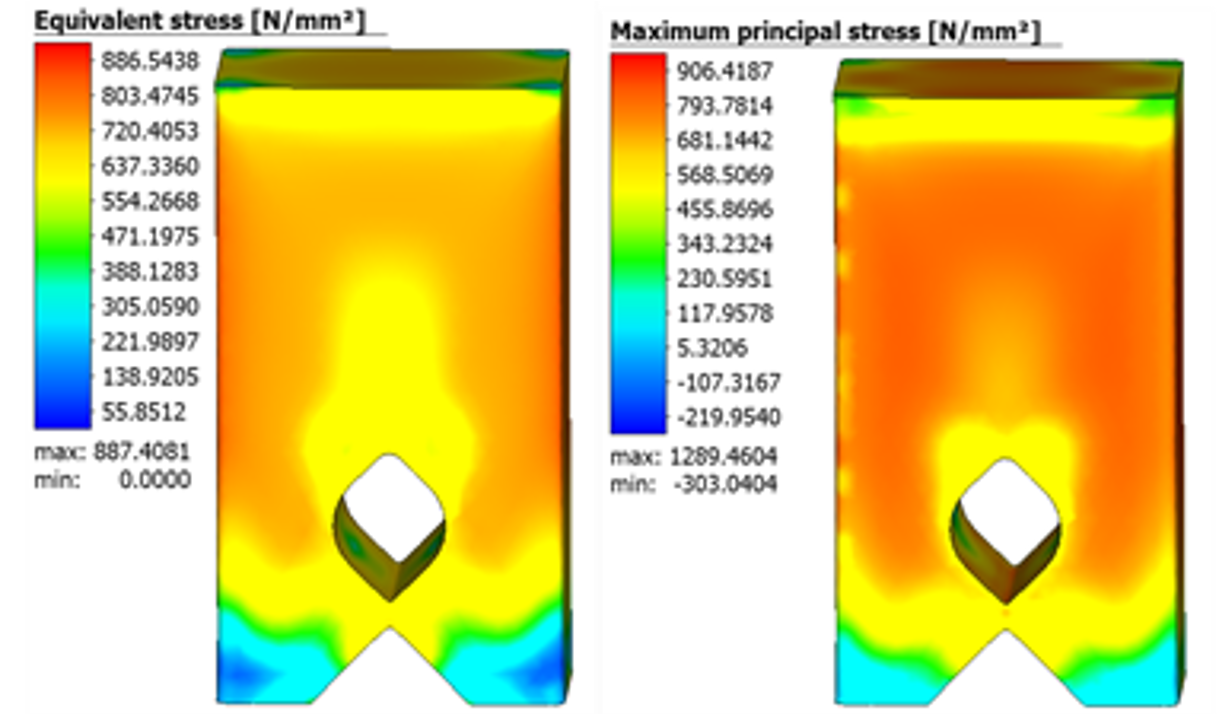}
	\caption{Exploring 3D Maximum Principal Stress and Equivalent Stress as potential crack indices.}
	\label{fig_crackidx}
\end{figure}

The specimen may crack during the build, end of build, or during post-processing. Since cracks are a mechanism to relieve stresses, the residual stresses at the end of build process is studied in this work. To predict failure of the coupon during or at the end of the build process requires a reliable failure parameter that can repeatably predict the failure location. Parameters like maximum principal stress and equivalent stress estimated at the end of build process, which are typically used in structural analysis to detect failure, were initially explored in the hope that they would accurately correlate with crack risk (propensity) and identify the location of crack. However, it was found, as shown in Fig.~\ref{fig_crackidx}, that neither of those two parameters indicate the location of crack seen during builds with precision. 

\begin{figure*} [ht!]
	\centering
	\includegraphics[width=\linewidth]{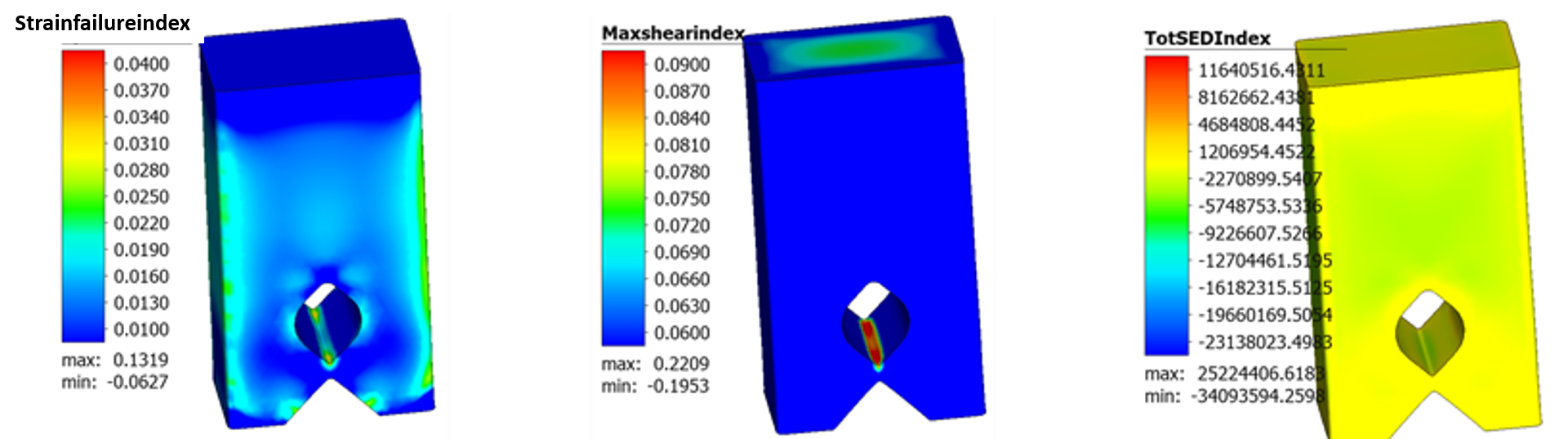}
	\caption{Exploring 3 additional crack indices. MSSI best captures the region of crack with high precision.}
	\label{fig_3idx}
\end{figure*}

As seen in the Fig.~\ref{fig_crackidx}, both parameters seem to indicate that a crack could initiate from almost any location in the coupon, whereas it is known experimentally that the crack always initiates from the bottom of the cooling channel feature. This required a more detailed exploration of parameters for crack index. As mentioned earlier, in this work, we explored three failure parameters described in~\cite{bao2004comparative}: SFI, MSSI, and TSI that are estimated as follows:
\begin{align}
    SFI &= \frac{\tau \cdot \bar{\epsilon}}{{\epsilon}^{}_{UTS}} \nonumber\\
    MSSI &= \frac{\tau \cdot {(\epsilon^{}_1 - \epsilon^{}_3) }}{{\epsilon}^{}_{UTS}}\nonumber\\
    TSI &= \frac{\tau \cdot {(\sigma^{}_i \cdot \epsilon^{}_i) }}{({\sigma}^{}_{UTS} \cdot {\epsilon}^{}_{UTS})} \nonumber
\end{align}

where $\sigma^{}_i$ are the principal stresses, $\epsilon^{}_i$ are the principal strains,  ${\bar{\epsilon}}$ is the effective plastic strain, ${\epsilon}^{}_{UTS}$ is the strain at the ultimate tensile strength, and $\tau$ is the triaxiality factor given as,
$$ \tau = \frac{{\sigma}^{}_{mean}}{{\sigma}^{}_{vonMises}} $$
Figure~\ref{fig_3idx} shows the results of estimating each of the three criteria from the output of an inherent strain based additive process simulation applied to a coupon that is known to crack at the bottom of the cooling channel (i.e., the “No-Go” coupon). After adjusting the thresholds, it can be seen that the second criterion, i.e., the MSSI criterion indicates a high value for the observed crack location with good precision. As a result, MSSI was chosen as the crack index for the outcomes described in this paper.
\subsection{Training Data Generation}\label{sec_trainingdata}

\begin{figure*} [ht!]
	\begin{subfigure}[t]{0.2\linewidth}
		\centering
		\raisebox{0.25\height}{\includegraphics[width=0.7\linewidth]{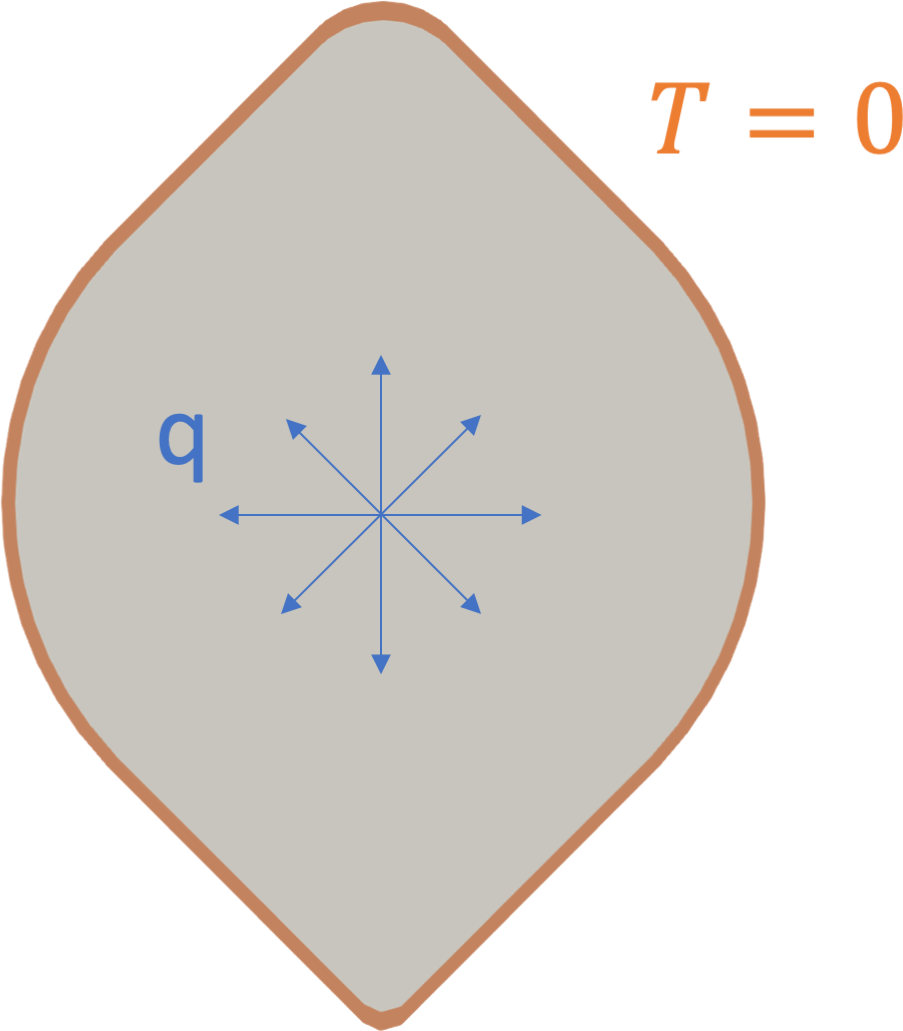}}
		\captionsetup{justification=centering}
		\caption{Thermal problem}
	\end{subfigure}%
	\begin{subfigure}[t]{0.8\linewidth}
		\centering
		\includegraphics[width=0.95\linewidth]{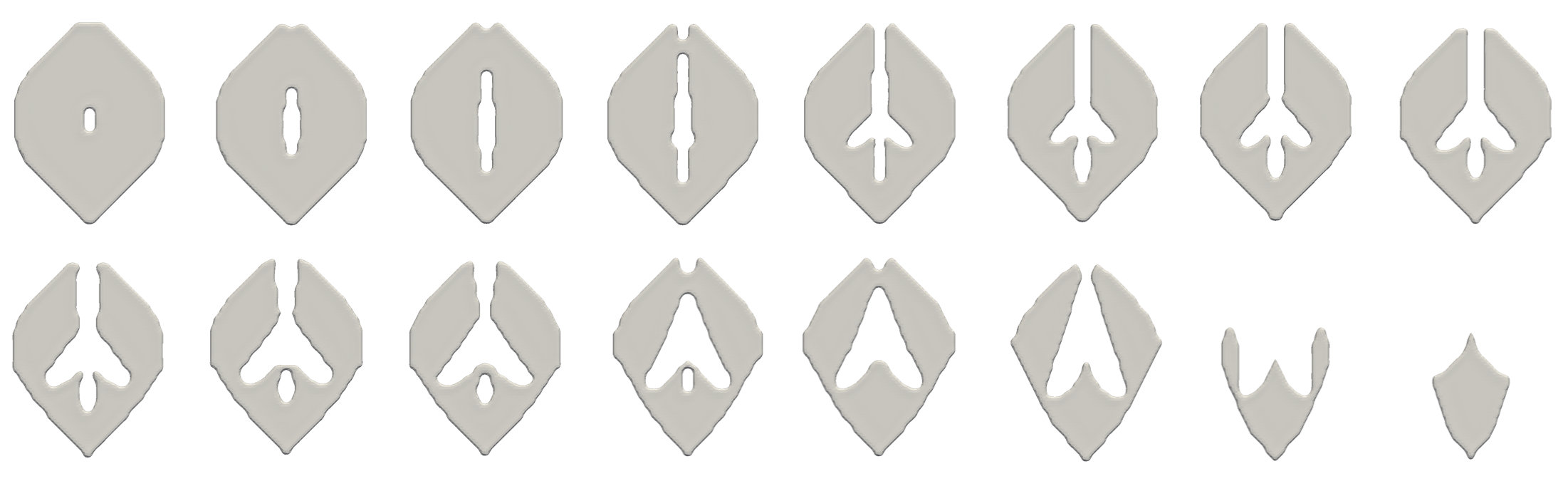}
		\captionsetup{justification=centering}
		\caption{Training data at various volume fractions for the symmetric thermal problem with self-supporting constraint}
	\end{subfigure}%
	\\ 
	\begin{subfigure}[t]{0.2\linewidth}
		\centering
			\raisebox{-0.3in}{\includegraphics[width=0.75\linewidth]{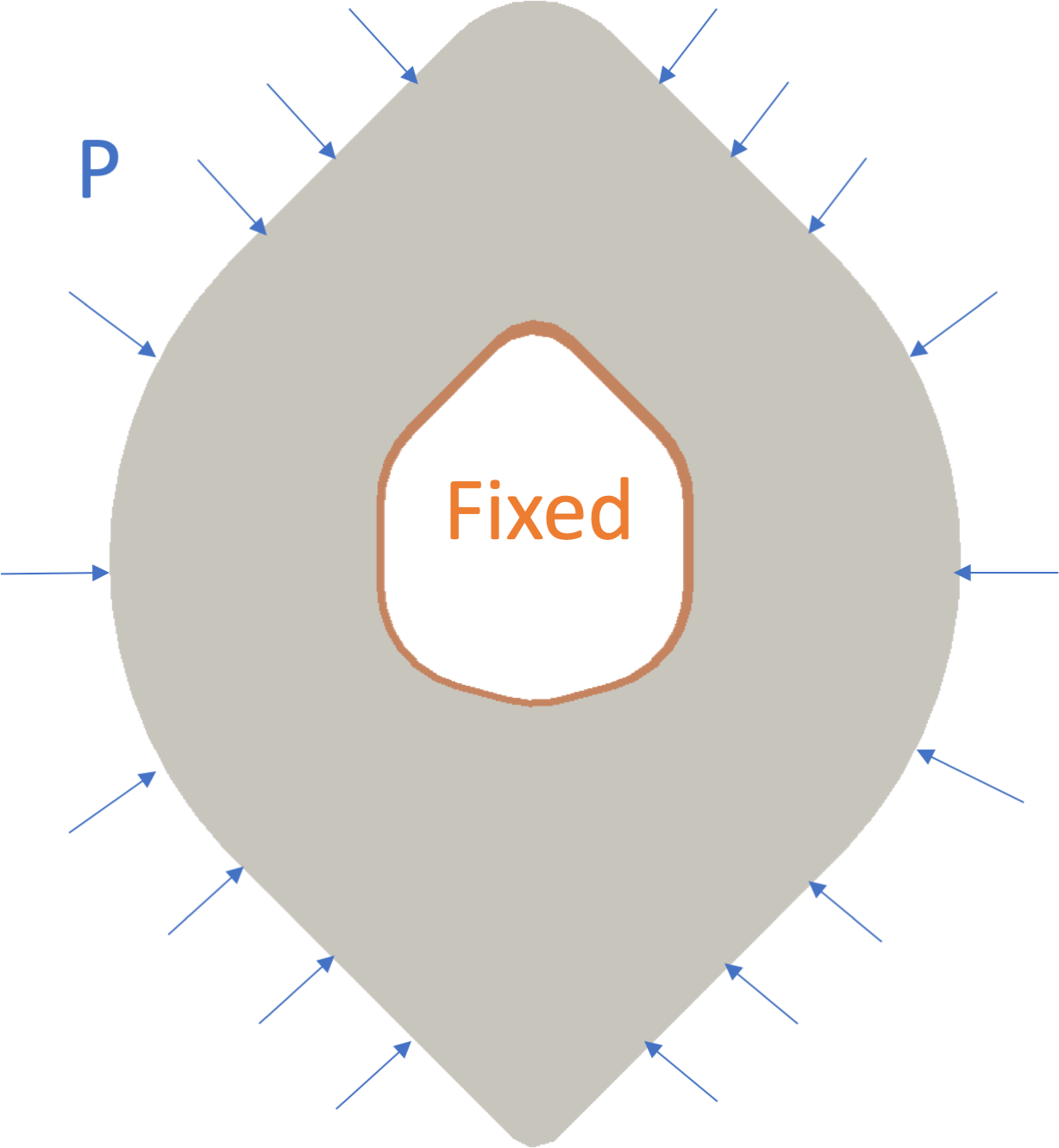}}\hspace{10pt}
		\captionsetup{justification=centering}
		\caption{Hydrostatic pressure problem}
	\end{subfigure}%
	\begin{subfigure}[t]{0.8\linewidth}
		\centering
		\raisebox{-0.3in}	{\includegraphics[width=0.75\linewidth]{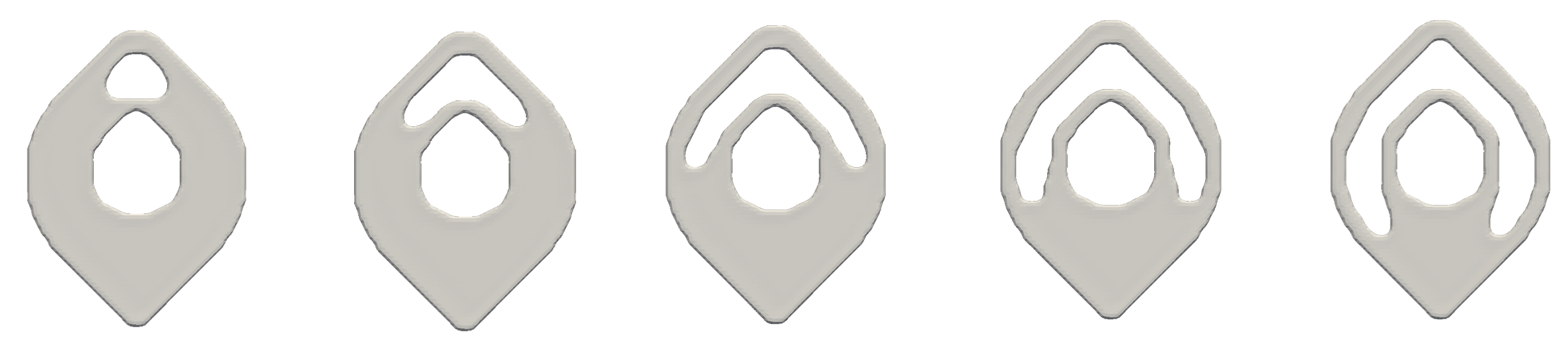}}
		\captionsetup{justification=centering}
		\caption{Training data at various volume fractions the hydrostatic pressure problem}
	\end{subfigure}%
	\\
	\begin{subfigure}[t]{0.2\linewidth}
		\centering
		{\includegraphics[width=0.7\linewidth]{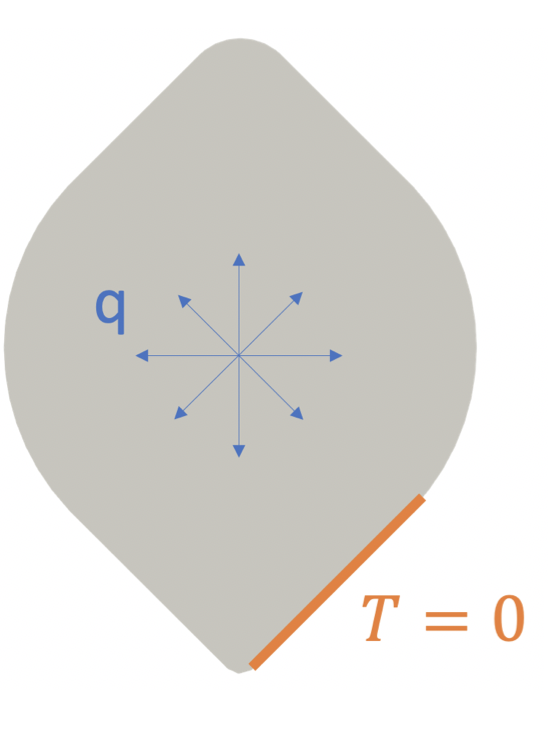}}
		\captionsetup{justification=centering}
		\caption{Asymmetric thermal problem}
	\end{subfigure}%
	\begin{subfigure}[t]{0.8\linewidth}
		\centering
			\raisebox{0.3in}{\includegraphics[width=0.95\linewidth]{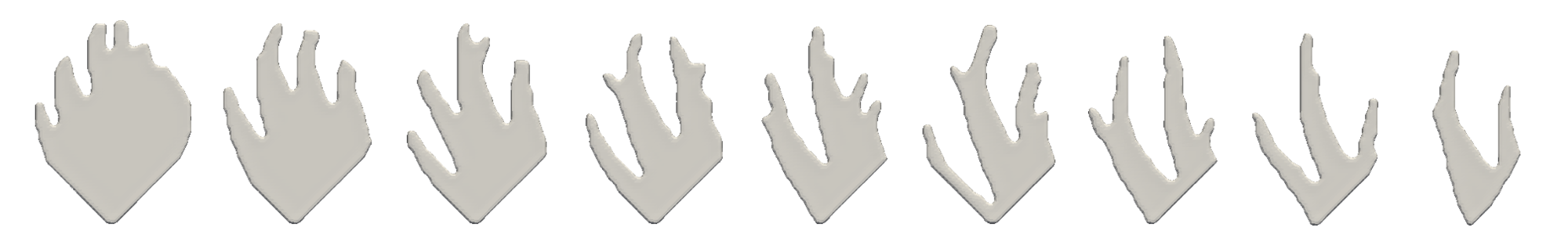}}
		\captionsetup{justification=centering}
		\caption{Training data at various volume fractions for the asymmetric thermal problem}
	\end{subfigure}%
	\\
	\begin{subfigure}[t]{0.2\linewidth}
		\centering
		\includegraphics[width=0.7\linewidth]{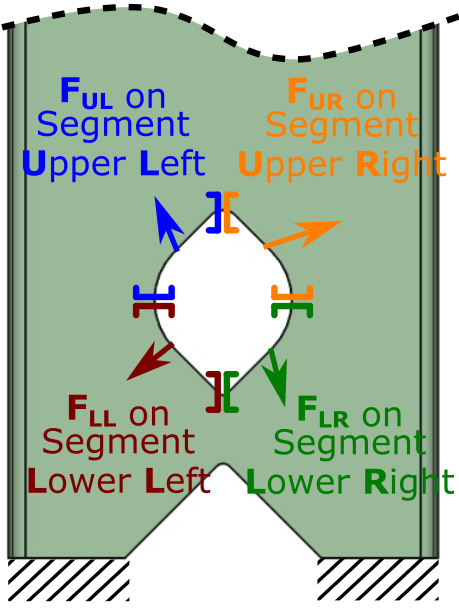}
		\captionsetup{justification=centering}
		\caption{Four-segment loading problem}
	\end{subfigure}%
	\begin{subfigure}[t]{0.8\linewidth}
		\centering
		\includegraphics[width=0.75\linewidth]{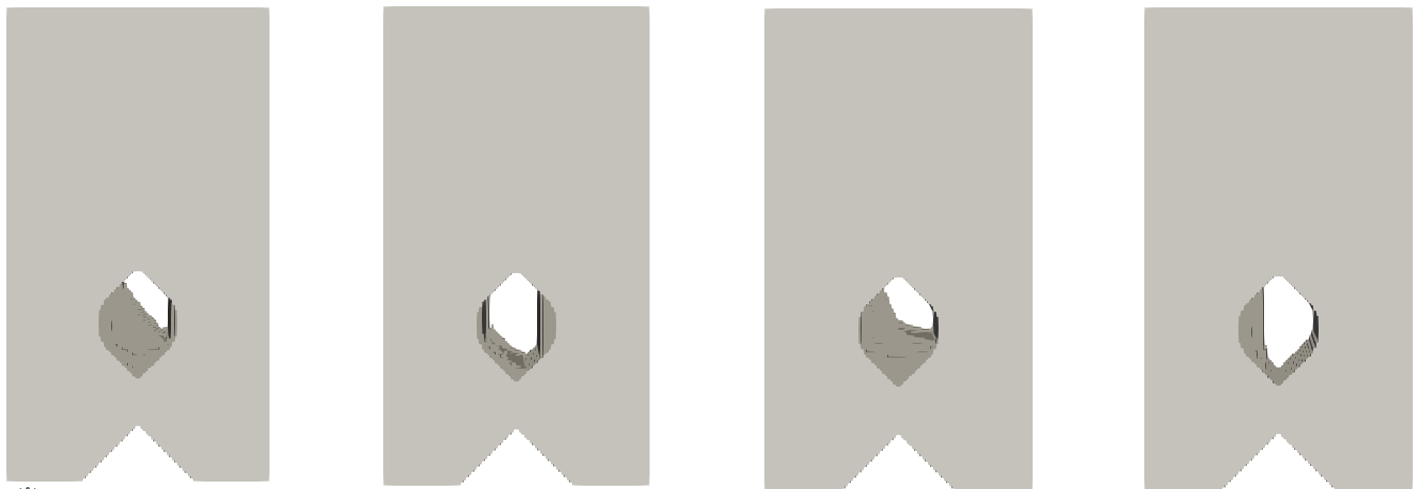}
		\captionsetup{justification=centering}
		\caption{Training data generated with various values of the TO parameters summarized in the Table 1.}
	\end{subfigure}%
	\caption{Training data was generated for different boundary conditions, volume fractions, and design and manufacturing constraints.} \label{fig_trainingData}
\end{figure*}

Since the surrogate  model will be used within the TO loop, it is essential that sufficient training data is provided  by  TO  under  different physics and boundary conditions, while considering relevant design and manufacturing constraints mentioned in Section~\ref{sec_problem}. Figures~\ref{fig_trainingData}(a) and (b) illustrate the thermal conduction problem and a small sample of symmetric self-supporting designs of the channel to maximize thermal conductivity at different volume fractions. Figure ~\ref{fig_trainingData}c shows the hydrostatic pressure problem, where the initial design domain is slightly modified to include a self-supporting through-cut channel. The optimized designs for the hydrostatic pressure problem are shown in Fig.~\ref{fig_trainingData}(d), where the structural compliance is minimized.  Figures~\ref{fig_trainingData}(e) and (f) illustrate an asymmetrical thermal problem and the corresponding optimized designs at different volume fractions, respectively. In addition to the constant thermal and pressure loading, a set of 3D TO problems has been considered featuring varying distribution of surface loading on the surface of the channels. To be specific, the channel surface of a baseline/initial design is divided into 4 segments (see Fig.~\ref{fig_trainingData}(g)) by the vertical and horizontal symmetrical planes of the channel's cross-section shape. The objective of the TO problems is structural compliance minimization. In order to introduce sufficient diversities into the training data set, the surface loading applied on each of these segments can vary independently. In addition, two levels of volume fraction constraints are considered to further diversify the generated geometries. The additive manufacturability filter developed by Langelaar~\cite{langelaar_additive_2017} has been applied to ensure the generated geometries are free from self-support issues in additive fabrication. This maintains the relevancy of the training data for the surrogate model. 540 geometries were generated through this approach. A small set of the generated samples are shown in Fig.~\ref{fig_trainingData}(h). 

\subsection{Training Data Selection}\label{sec_trainingSelect}

\begin{figure*} [ht!]
	\centering
	\includegraphics[width=0.9\linewidth]{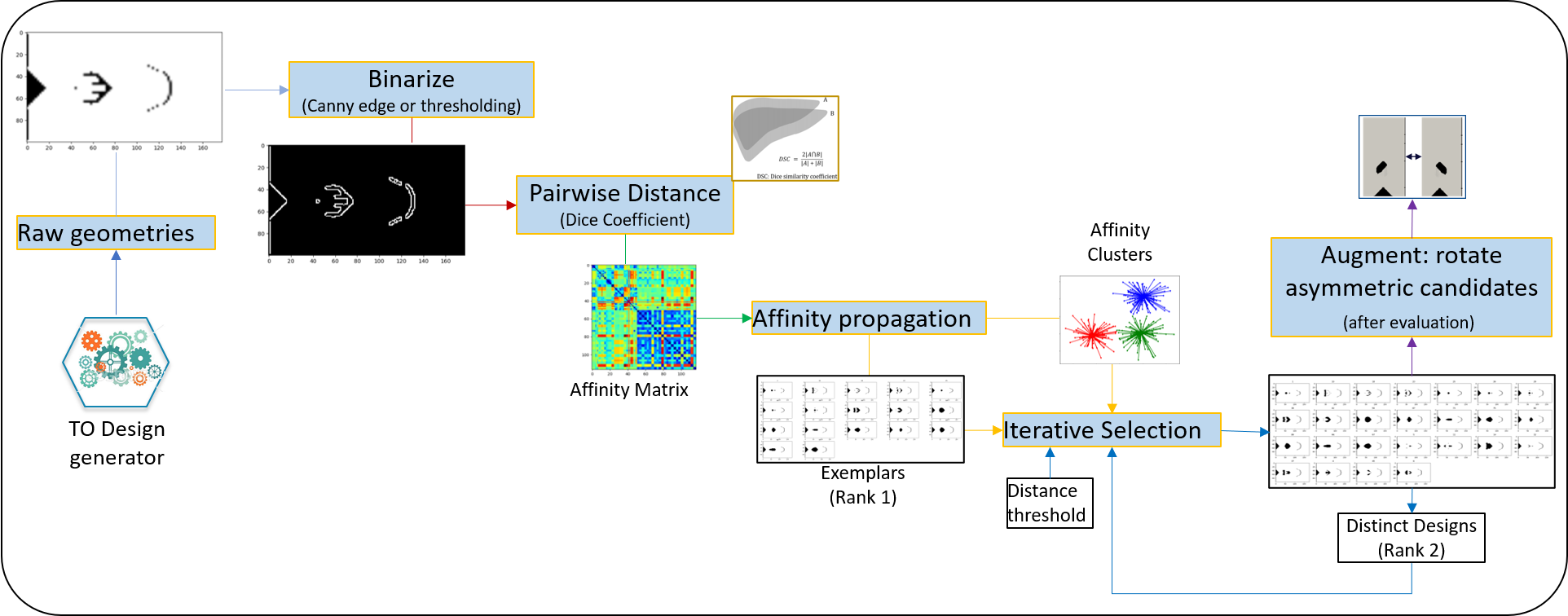}
	\caption{A workflow for selection of maximally diverse subset of design samples from a given set.}
	\label{fig_data_sel}
\end{figure*}

The accuracy and generalizability of the surrogate model will benefit from training the surrogate on a larger, as well as a more diverse, set of channel design samples. This requirement is additionally critical due to the time-complexity of evaluating a single sample using the high-fidelity simulator. Since the relationship of the parameters that were varied for training data generation (for e.g., loading, boundary conditions etc.) to the geometry of the design is complex, a purely parametric approach might still likely produce design geometries that are only incrementally different. To address this issue, an additional sample selection module was developed, that helps down-select from a larger set of samples, generated freely by the design generator, to a subset that are most diverse with respect to geometric features. This larger, and more diverse set was used to train the next surrogate. Figure~\ref{fig_data_sel} shows a workflow that was designed to select a maximally diverse set of design samples, where the input is a set of 3D samples generated by the design generator. Given that the designs are largely symmetric along the y-axis, we make use of the central xz-slice of the 3D volume to represent a sample, so that design similarity (or diversity) measured across a pair of 2D slices can be expected to apply without loss of generality to the corresponding pair of 3D samples as well. These input samples are non-binary based on the interpolation step performed to upsample the low resolution simulation runs to high resolution. Although the new set of samples (540 in number) are fairly diverse, they also include samples that are only incrementally different from each other. The objective is to weed out any such samples that are already represented by some other sample in the set, thereby creating a maximally diverse subset of samples necessary for training a surrogate that is robust and generalizable in the design domain. In some sense, the size of this subset can be thought of as the true rank of the generated design space. 

\begin{figure} [ht!]
	\centering
	\includegraphics[width=\linewidth]{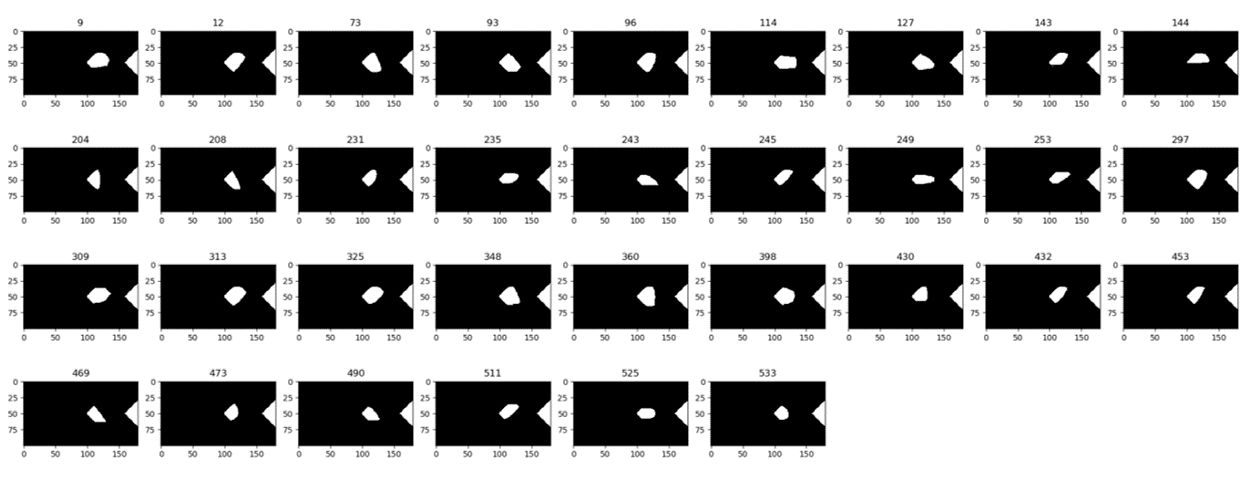}
	\caption{The 33, rank-1 exemplars using Affinity Propagation on 540 design samples.}
	\label{fig_exemplars}
\end{figure}

The grayscale geometry samples are initially subjected to a binarization step using either an edge detection algorithm like Canny edge detector~\cite{Canny_Edge} or by using an appropriate threshold. Now that all the 540, 2D slices are binarized, the goal is to create a pairwise affinity or similarity matrix where matrix-entry (i, j) measures the similarity between samples ‘i’ and ‘j.’ Multiple distance metrics like Euclidean or Hausdorff can be used, but we make use of the Dice Coefficient~\cite{Dice,Srensen1948AMO} to identify the extent of overlap between pairs of design samples. The Dice coefficient is ideal because the standard viewpoint for all design samples is known and there is thus no need to consider complexity related to rotational variance when comparing 2 samples. Another advantage of using the Dice coefficient is that its range is naturally limited between 0 and 1, thus automatically normalizing the affinity matrix for all the design samples. 
The affinity matrix for the 540 samples is now processed to identify clusters of samples within the set, that share affinity values that are close to each other. An approach ideally suited for this is the Affinity Propagation algorithm~\cite{Frey2007ClusteringBP}. Given an affinity matrix, this algorithm uses message-passing between the samples to converge to a state that allows the inference of the number of affinity clusters and the choice of the specific samples (called the exemplars) that are considered to be representatives for a given affinity cluster. Unlike other clustering algorithms like k-means, Affinity Propagation does not require the user to specify the number of clusters; rather it is assumed that the affinity matrix implies the existence of these clusters and we arrive at those when the algorithm converges. Also, clustering approaches like k-means generate synthetic prototypes as cluster representatives and thus the cluster centers are not themselves samples; in contrast, affinity propagation picks actual samples as exemplars to represent a cluster center. Because they are actual samples, the set of exemplars produced by the approach can be viewed as a maximally, diverse subset of samples for the entire set. We term these as ‘rank 1’ exemplars. Figure~\ref{fig_exemplars} shows the 33 rank-1 exemplars that were inferred from the original set of 540 samples. The figure shows how characteristics like shape, extent, and orientation of the holes play a part in defining diversity across the samples. To further augment the set of exemplars in order to increase the number of training samples, we iteratively identify additional samples from the remaining set such that they are also maximally diverse – we term the $k^{th}$ such iteration as the rank-$k$ exemplar set. 

\begin{figure} [ht!]
	\centering
	\includegraphics[width=\linewidth]{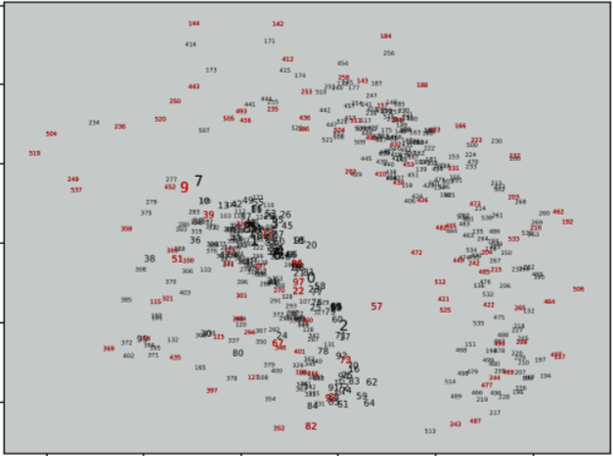}
	\caption{Multidimensional Scaling plot for all 540 designs, with red points showing the maximally diverse subset of designs that were chosen to be in the training data.}
	\label{fig_mds}
\end{figure}

In order to generate the augmented sets, in each iteration we extract one additional sample from each cluster, which is also farthest (in terms of its affinity value) from the exemplar samples for that cluster and the other exemplars in the cluster selected in previous iterations. We also make sure the chosen exemplar is beyond a certain distance-threshold from the nearest exemplar to preserve diversity. The iterations stop when there as no more new exemplars that meet the criteria. To illustrate the extent to which the chosen samples are both maximal and diverse, we plot all the samples in a 2D space constructed using multidimensional scaling (MDS) as shown in Fig.~\ref{fig_mds}. 

The red points indicate the 106 samples that were chosen as maximally diverse by our approach. As can be seen, the approach robustly picks out outlier samples as well as keeps these samples as far away from each other in this 2D Cartesian space as possible. In general, no portion of the overall space of designs is underrepresented, but no region is distinctly over-represented either, which is the desirable outcome we were targeting. Finally, given that the newly generated designs are not symmetric about the x-axis, we augment the set of 106 designs with versions of the designs that are rotated 180 degrees along the z-axis, so that the surrogate can deal with both versions of the samples. This leads to the selection of a training sample set of 244 samples, which was used to train the surrogate model.

\subsection{Training Data Evaluation}\label{sec_trainingEval}
\begin{figure} [ht!]
	\centering
	\includegraphics[width=\linewidth]{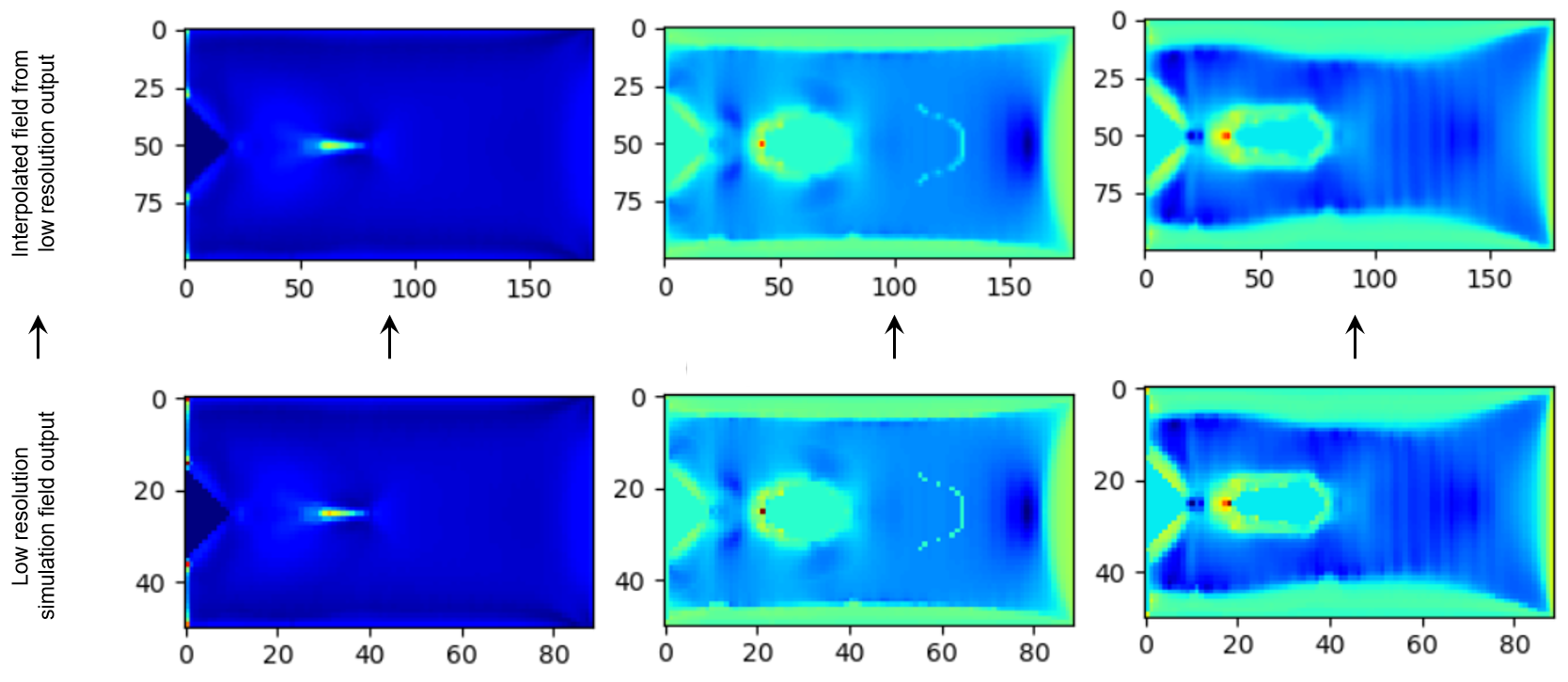}
	\caption{Comparison of the low resolution field produced by simulation with the corresponding high resolution field constructed using interpolation (3 examples).}
	\label{fig_interp2}
\end{figure}

Training data evaluation involves simulating the additive process for each training sample (i.e., channel design) and estimating the crack index field across the 3D coupon volume. As mentioned earlier, the inherent strain based methodology is used to simulate the build process. The part is placed at the center of the build plate and the build plate is constrained at four locations in all degrees of freedom to simulate the bolting boundary condition. The calibrated inherent strains mentioned in Section~\ref{subsec_mssi} are applied and each layer is deposited sequentially from bottom to top. The build process is simulated using commercially available Simufact Additive 2020 FP1 software package~\cite{simufact}. Given that we need to evaluate 244 input geometries to train the surrogate, we approximated the evaluation process, to make it time-efficient, as follows: the sample evaluation using the simulation is done at a much lower resolution (bringing down the run-time of each sample to only 4 hours), and then trilinear interpolation is used to reconstruct the crack index field produced by the coarse simulation into the equivalent field at the desired resolution. Figure~\ref{fig_interp2} shows 3 samples for which the low resolution crack index field produced by the simulator is compared with the equivalent high resolution field produced using interpolation. It shows that the interpolation while it visibly leads to smoothing, still retains the critical spatial characteristics of the signal.

\begin{figure} [ht!]
	\centering
	\includegraphics[width=\linewidth]{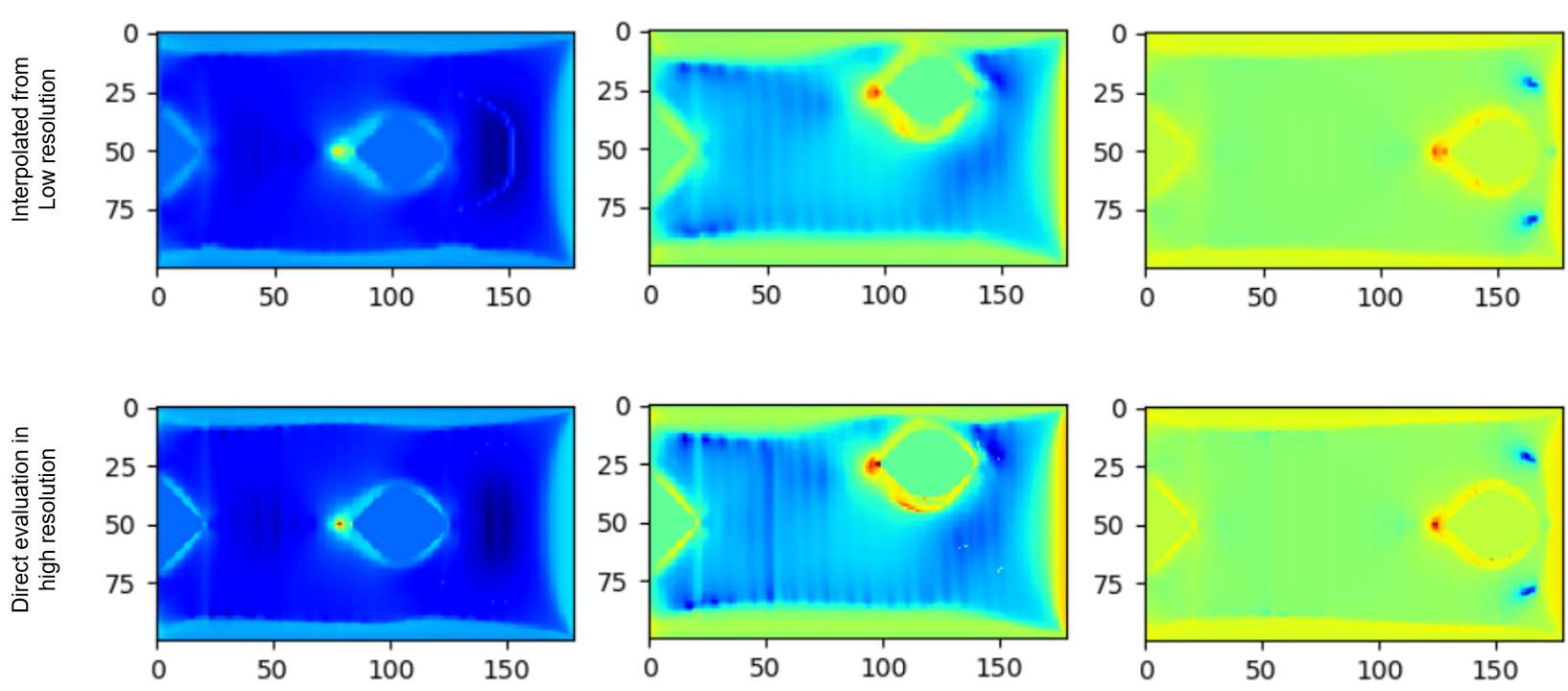}
	\caption{Comparison of the high resolution field produced directly by simulation with the corresponding high resolution field constructed using interpolation from a low resolution simulation of the same field (3 examples).}
	\label{fig_interp1}
\end{figure}

As an additional check on this approach, 12 geometries were simulated, at both low and high resolution to validate the outcomes. This is shown in Fig.~\ref{fig_interp1} and it shows that the interpolation retains the original signal with a reasonable fidelity. Since we had the high resolution ground truth for these 12 geometries from simulation, we further estimated MRE to compare the two signals and estimated it to be 22\% (or 78\% accuracy). For the 244 geometries generated for surrogate development, all samples were evaluated using the process simulator at low resolution (which required ~4 hours of high performance computing run time per geometry). The crack index fields produced by the simulator were then upsampled by interpolating to the higher resolution field for use in surrogate training.

\subsection{3D Surrogate Model using Deep Learning: overview} \label{sec_surrogateOverview}
\begin{figure} [ht!]
	\centering
	\includegraphics[width=0.95\linewidth]{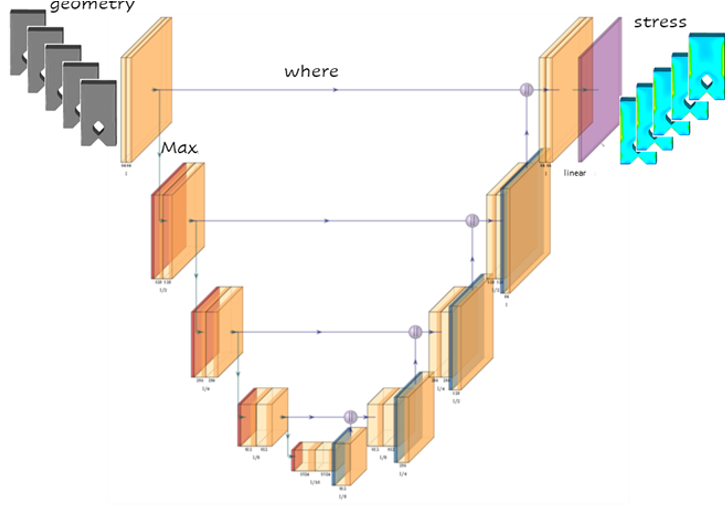}
	\caption{A 3D U-Net architecture for dense regression of the crack index field (from \cite{Iyer2021AttentionBased3N}).}
	\label{fig_baseline}
\end{figure}

\begin{figure*} [ht!]
	\centering
	\includegraphics[width=0.68\linewidth]{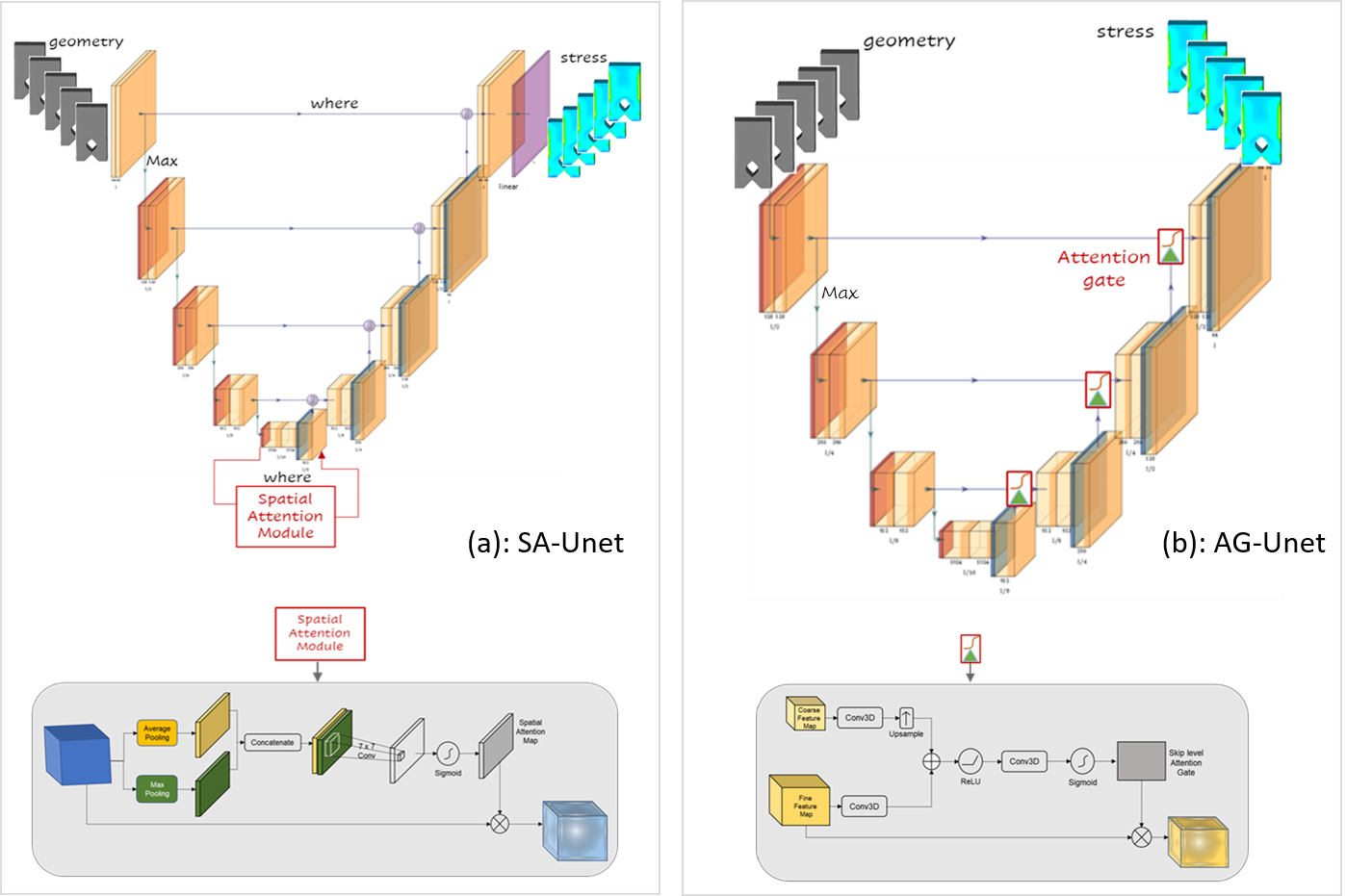}
	\caption{Two alternate Attention mechanisms (\cite{Guo_SAUnet,Schlemper2019AttentionGN}) to augment the U-Net architecture as explored in \cite{Iyer2021AttentionBased3N}.}
	\label{fig_attn_arch}
\end{figure*}

In recent work~\cite{Iyer2021AttentionBased3N}, we presented early outcomes leveraging deep convolutional neural networks (DCNN) as high fidelity and time-efficient surrogates of the 3D crack index field. We applied the U-Net architecture~\cite{Olaf_Unet} as the baseline, expanding the standard application of this architecture for 2D segmentation to the estimation of the full 3D, continuous valued, crack index field, as shown in Fig.~\ref{fig_baseline}. The paper illustrated the primary challenge faced by the standard U-Net architecture with L2-loss arising from sparsity in critical values of the crack index - part regions with high values of the crack index are often in a much smaller minority of the overall volume of the dataset used to train the surrogate. And we showed examples where, the use of standard metrics of loss like L2-loss can lead to a surrogate that only learns to reliably predict in regions where the crack index values are from the likely values of the overall distribution, but ignores or poorly models the rarer high values of crack index, which are critical to the problem at hand. 
In response, our paper demonstrated how the the idea of Attention, as inspired from cognitive attention as seen in humans, and predominantly applied in the natural language processing community~\cite{Bahdanau2015NeuralMT,Vaswani_Attention}, leads to promising outcomes for the crack index prediction problem. We explored how augmenting the U-Net architecture with two alternative attention mechanisms (see Fig.~\ref{fig_attn_arch}) helps address the issue as well as improve the overall accuracy of estimation. 

\begin{figure*} [ht!]
	\centering
	\includegraphics[width=0.64\linewidth]{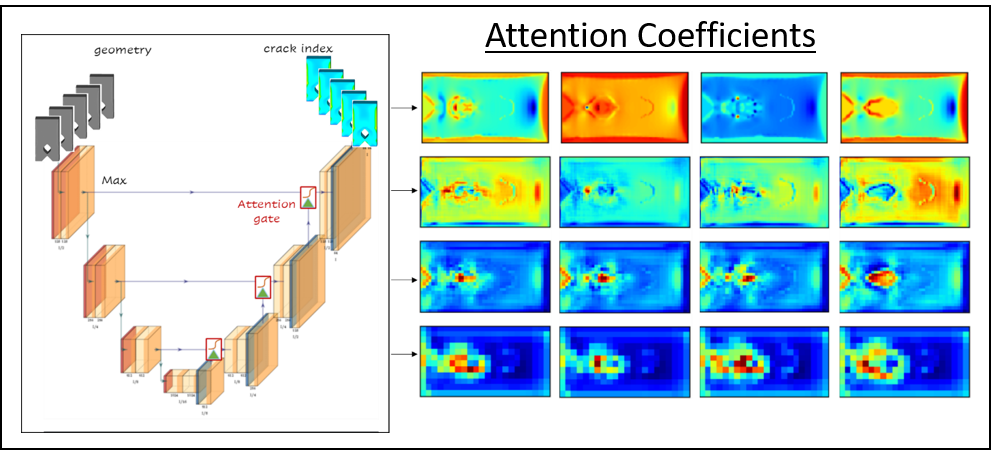}
	\caption{Multi-scale features for the crack index captured by Attention coefficients in the skip connections of AG-Unet from \cite{Iyer2021AttentionBased3N}.}
	\label{fig_attn_coeffs}
\end{figure*}

\begin{figure*} [!h]
	\centering
	\includegraphics[width=0.5\linewidth]{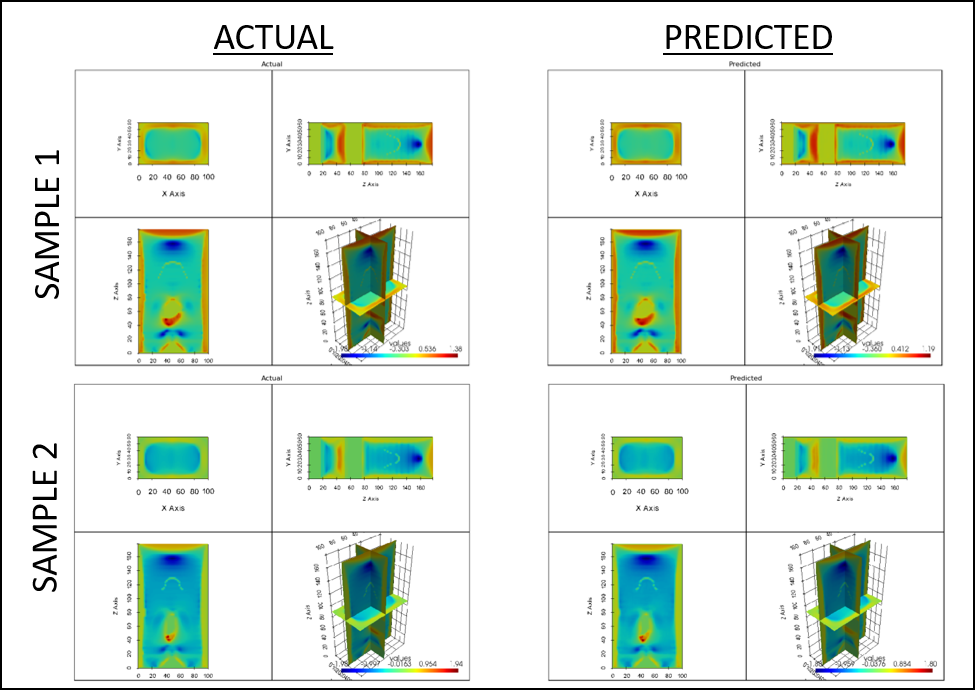}
	\caption{Predictions of the AG-Unet surrogate for 2 test samples and comparison with the ground truth crack index field.}
	\label{fig_outcomes}
\end{figure*}

One variant of the Attention mechanism, shown in Fig.~\ref{fig_attn_arch}(a), is inspired from~\cite{Guo_SAUnet} where the spatial attention map is computed on the bottleneck connecting the Encoder of the U-Net to the Decoder: the gating operation for this Spatial Attention Gate entails applying average-pooling and max-pooling operations on the bottleneck features, concatenating them and then using a single $7 \times 7$ Sigmoid-activated, convolution kernel to construct the spatial attention map. The attention map acts as spatial weights to help focus on the regions most relevant for accurate estimation of the crack index, by emphasizing or suppressing appropriate features from the bottleneck. The second approach, shown in Fig.~\ref{fig_attn_arch}(b), is inspired from~\cite{Schlemper2019AttentionGN} in which an additive attention gating mechanism is encoded in the skip connections, allowing for attention-based coefficients to be learned specific to sub-regions in the image at different spatial scales. In other words, the gating signal helps amplify critical, task-specific, and spatial features in the input at multiple scales, that is already encoded in the skip connections in a U-Net. While both the Attention-based instantiations were shown to improve mean voxelwise prediction accuracy compared to the baseline U-Net model, the AG-Unet model was shown to capture salient aspects of the crack index distribution relatively better as well as produce more reliable estimates of the rare, but high values of the crack index which is critical since these regions of the part are the ones most prone to cracking. The primary reason for this was further illustrated by looking at the attention coefficients that are created by the AG-Unet during the prediction task.

Figure~\ref{fig_attn_coeffs} shows the attention coefficients generated at each of the 3 skip connections of the AG-Unet architecture (rows 2, 3, and 4) for 4 design samples; the topmost row shows the actual crack index value. The figure shows more clearly how these coefficients spatially weight all the feature maps at each skip connection so as to emphasize portions of the feature-maps in regions of the part where dominant geometric features are visible at that scale. Finally, attention mechanisms can help regularize the learning process to construct the right semantic representation and put the network parameters to use in learning the right function that is also semantically aligned with the task at hand. More specifically, the presence of geometric features semantically signal the existence of interesting behaviour of the crack index near those features; attention mechanisms help reinforce the need for the network to learn those crack index values better. 

In this paper, we explore the same 3 architectures, but trained on a larger and more diverse training dataset compared to what was use to generate the outcomes in~\cite{Iyer2021AttentionBased3N}. 

\subsection{3D Surrogate Model using Deep Learning: outcomes}\label{sec_surrogateOutcome}
As described in the previous section, we consider the 3D U-Net architecture and its variants for voxel-wise regression that makes use of {\it MaxPooling} for feature abstraction. The high computational cost of training a 3D U-Net is addressed by conducting the training on an \textsf{NVIDIA}$^{\small{\textregistered}}$ DGX machine configured with 8, P100 GPUs. Table~\ref{tab0} shows some details of the training regimen, that was used to train all 3 surrogates. 

\begin{table} [!h]
	\caption{Training regimen characteristics for the surrogates}
	\begin{center}
		\begin{tabular}{  p{38mm}  | p{43mm}  }
			Training Parameter & Description  \\ \hline
		Architecture &  {3D U-Net, $3 \times 3$, ReLU,\newline 1-32-64-128-256-128-64-32-1}\\
		
		Feature Abstraction &  {Max Pooling}\\
		
		Output Layer Activation & {Linear}\\
		
		Loss Function & {L2/MSE}\\
		
		Target & {Raw crack index values}\\
		
		Optimizer & {ADAM, init-lr=1e-4}\\
		
		Number of samples &  {240}\\
		
		Batch Normalization &  {No}\\
		
		Dropout &  {No}\\
		
		Input sample & {$100 \times 60 \times 178$ volume}\\
		
		Train/Test split & {192 / 48 samples}\\
		
		Batch Size per GPU & {8}\\
		
		\#epochs & {150, with early stopping check}\\
		
		Early Stopping & val-loss,\newline min-delta=1e-9,\newline patience=5\\
		
		Initialization & {Glorot-uniform}\\
		
		Compute & {Parallel 4-GPU}\\
		
		\end{tabular}
	\end{center}
	
	\label{tab0}
\end{table}

\begin{table} [!h]
	\centering
	\caption{Performance comparison}
	\tabulinesep=0.5mm
	\begin{tabu}[t!]{p{53mm}cc}
		\hline \hline 
		Surrogate Description & \%MRE& \%Acc.\\  
		\hline 
		Standard 3D U-Net for regression &  14.13 &  85.8\\
		U-Net w/ Spatial Attention  &  13.75 & 86.2\\
		U-Net w/ Attention Gates on skip & 13.40 & 86.6\\
		\hline
	\end{tabu}
	\label{tab1}
\end{table}

\begin{figure} [!hb] \centering
	\begin{subfigure}[t]{0.4\linewidth}
		\centering
		\includegraphics[width=0.75\linewidth]{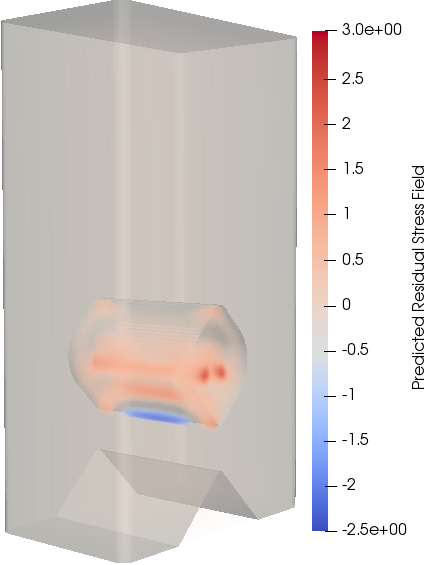}
		\caption{}
	\end{subfigure}%
	\begin{subfigure}[t]{0.4\linewidth}
		\centering
		\includegraphics[width=0.75\linewidth]{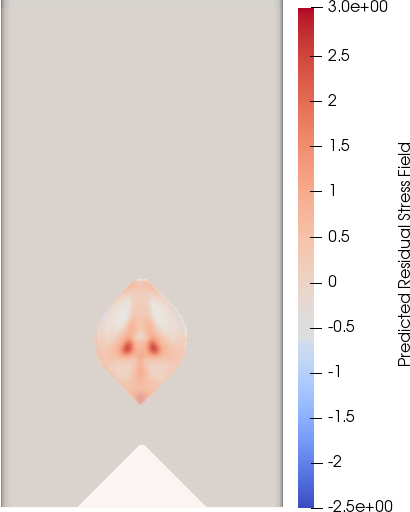}
		\caption{}
	\end{subfigure}%
	\\
	\begin{subfigure}[t]{0.4\linewidth}
		\centering
		\includegraphics[width=0.75\linewidth]{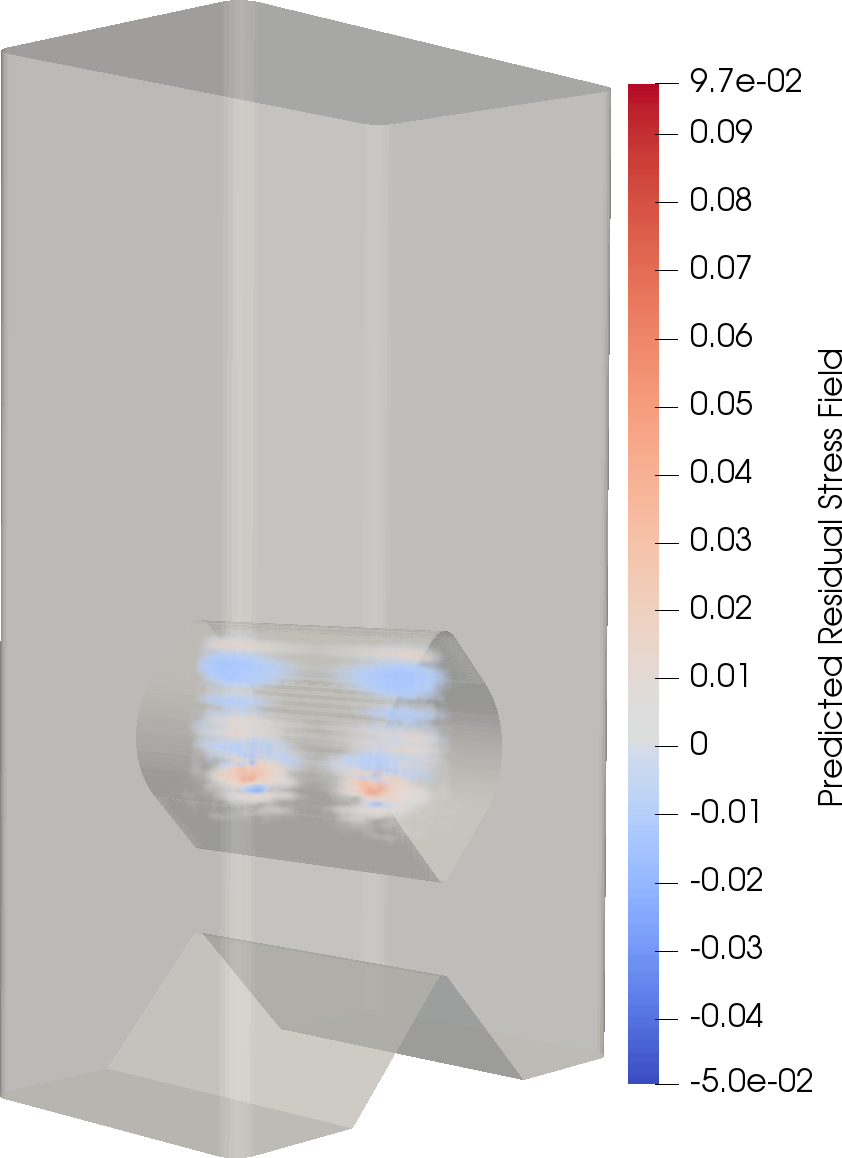}
		\caption{}
	\end{subfigure}%
	\begin{subfigure}[t]{0.4\linewidth}
		\centering
		\includegraphics[width=0.75\linewidth]{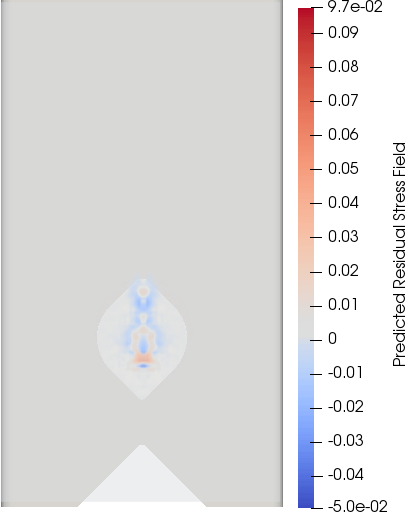}
		\caption{}
	\end{subfigure}%
	\caption{MSSI field inside the channel (top) and its gradient field with respect to maximum value (bottom).} \label{fig_MSSIinChannel}
\end{figure}

\begin{figure} [!h]
	\centering
	\includegraphics[width=\linewidth]{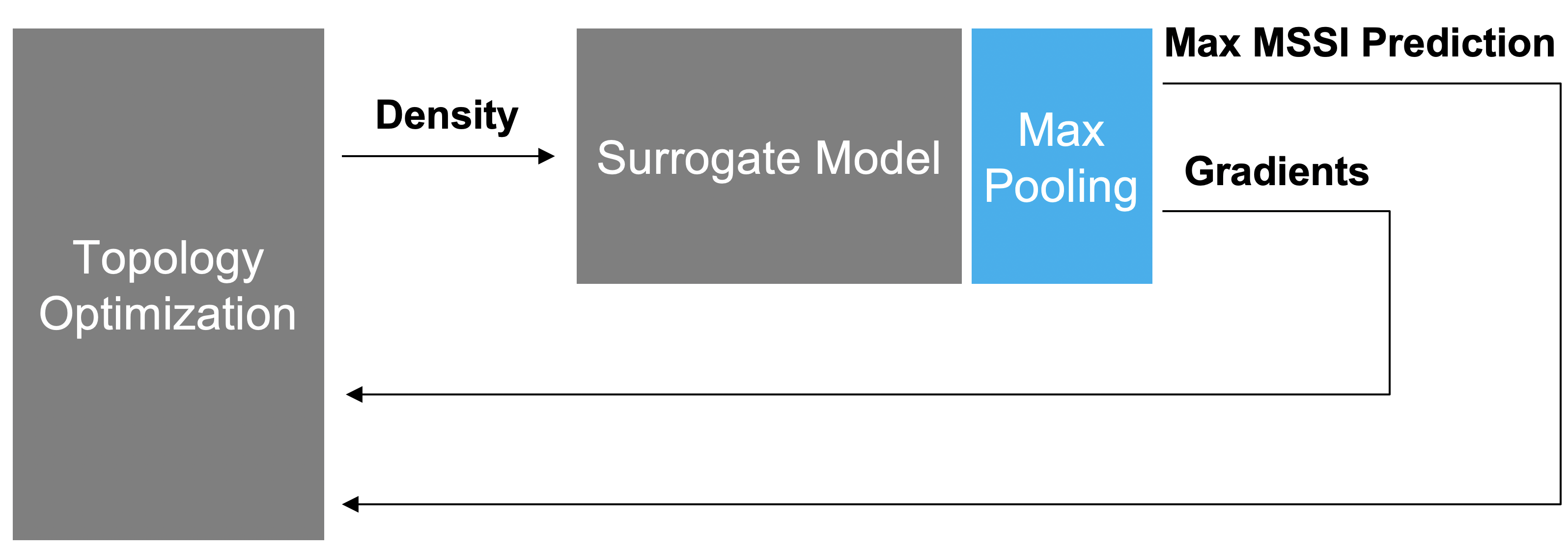}
	\caption{MSSI-aware TO with NN-based surrogate model. The MSSI sensitivity field is computed using automatic differentiation.}
	\label{fig_TOwithSurrogate}
\end{figure}

While L2-loss was used for training the surrogate and for measuring its performance, its values are in the units of the quantity being measured, in this case the crack index. Therefore, we make use of a normalized metric called {\it Mean Relative Error} (MRE) to additionally report performance of the model in terms of its mean voxelwise accuracy:
$$\text{MRE} = \frac{1}{n} \sum_{j=1}^{n} \frac {|y_j - \hat{y_j}|}{\epsilon+max(|y_j|,|\hat{y_j}|)}  $$
MRE captures the relative error rate in percent units when comparing deviation of predictions from ground truth, and thus expressed within the standard range between 0 and 1, which is desirable. Table~\ref{tab1} compares the relative performance of the 3 surrogates in terms of both MRE and Accuracy for each of the 3 surrogates.

The metrics in the table quantify the mean relative error (or accuracy) averaged across all voxels in the set of the 48 test samples. The numbers indicate that both baseline U-Net architectures with attention mechanisms perform better than the U-Net without any augmentation, with the AG-Unet providing the best performance. Figure~\ref{fig_outcomes} shows predictions of the AG-Unet for 2 of the test samples, illustrated through the 3 orthographic projections of the corresponding middle slices of the sample. 

\begin{figure*} [ht!]
	\centering
	\includegraphics[width=\linewidth]{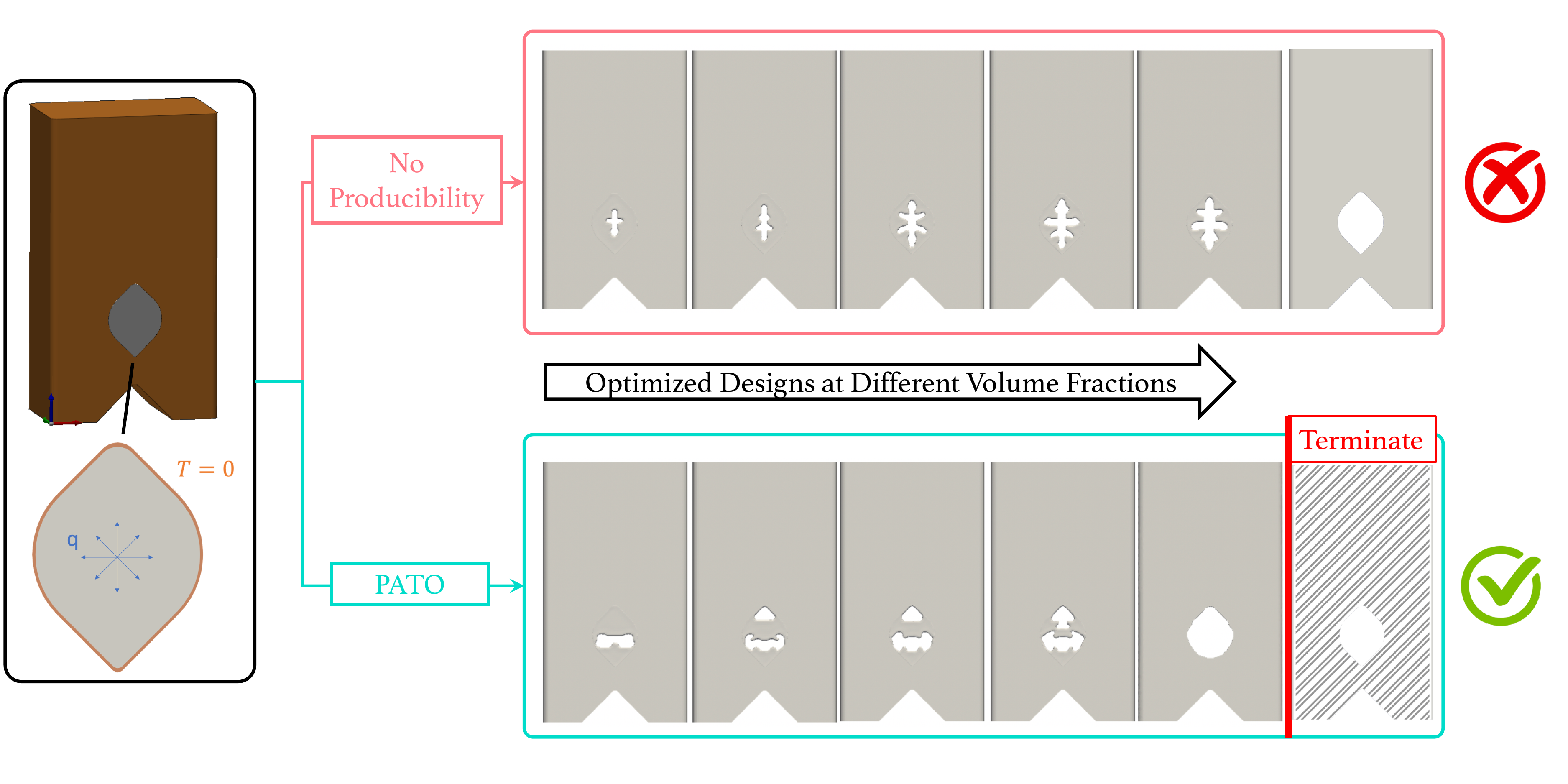}
	\caption{Optimized designs at different volume fraction for TO without considering producibility (top) and the proposed PATO (bottom).}
	\label{fig_MSSIConsVsUncVer1}
\end{figure*}

\subsection{TO Formulation and Sensitivity Analysis} \label{sec_TOformulate}

In this section, we will describe the multi-objective PATO formulation and the sensitivity analysis for computing the gradients by incorporating the AG-Unet surrogate model for MSSI prediction.  
Mathematically, the  crack-free TO problem considering the performance $\varphi$ and MSSI cracking index $\MSSImax$ is formulated as:
\begin{subequations}
	\label{eq_MSSIawareTO}
	\begin{align}
	\underset{\rho}{\text{minimize}} \quad & (1-w)\varphi(\bu) + w\MSSImax (\rho) \label{eq:optimization_base_objective}\\
	\text{s.t.} \quad
	& \bK(\rho)\bu(\rho) = \bff \label{eq:optimization_base_govnEq}\\
	& \frac{V(\rho)}{V_{\targ}} - 1 \le 0  \label{eq:optimization_volCons}\\
	& 0 \le \rho \le 1 \label{eq:density}\\
	& 0 \le w \le 1 \label{eq:weight}
	\end{align}
\end{subequations}

where $w$ and $\rho$ are the weighting factor and the pseudo-density design variable, respectively. $\bu$ denotes the state variable that satisfies the state equation of \eqref{eq:optimization_base_govnEq} solved using finite element analysis (FEA), where $\bK$ is the stiffness matrix and $\bff$ is the external load vector. Equation \eqref{eq:optimization_volCons} is the volume constraint, where $V_{target}$ is the target volume fraction.

Equation~\ref{eq_MSSIawareTO} can be expressed as minimization of the following Lagrangian:
\begin{align}
	\mathcal{L}(\rho)&:= (1-w)\varphi(\bu) + w\MSSImax (\rho)\nonumber \\
	&+~\lambda_1 (\dfrac{V(\rho)}{V_{\targ}} - 1) 
	+\boldsymbol\lambda_2^\mathrm{T} \Big(\bK(\rho)\bu(\rho) - \bff \Big).
	\label{eq_Lag}
\end{align}
Using the prime symbol $(\cdot)'$ to represent differentiation of a function with respect to the design variable $\rho$,
we obtain (via chain rule):
\begin{align}
	\mathcal{L}'(\rho) &= (1-w)\varphi'(\bu) + w\MSSImax' (\rho) \nonumber\\
	&+ \lambda_1\dfrac{{V'}(\rho)}{V_{\targ}} + \boldsymbol\lambda_2^\mathrm{T}
	\Big(\bK(\rho)\bu(\rho)\Big)',\\
	& = w\MSSImax' (\rho) + \Big((1-w)[ \dfrac{\partial \varphi}{\partial \bu} ]+ \boldsymbol\lambda_2^\mathrm{T}
	\bK(\rho)\Big)\bu'(\rho) \nonumber \\
	& +\lambda_1 \dfrac{{V}'(\rho)}{V_{\targ}} + \boldsymbol\lambda_2^\mathrm{T}
	\bK'(\rho)\bu(\rho). \label{eq_chain_rule}
\end{align}
Since computing $[\bu_\rho']$ is computationally prohibitive, $[\lambda_2]$ is chosen as the solution to the adjoint problem~\cite{Bendsoe2009topology,mirzendehdel2017hands} which reduces \eqref{eq_chain_rule} to:
\begin{align}
	&\mathcal{L}'(\rho) = w\MSSImax' (\rho) + \boldsymbol\lambda_2^\mathrm{T} \bK'(\rho)\bu(\rho) + \lambda_1 \dfrac{{V}'(\rho)}{V_{\targ}} 
	, \label{eq_Lag_prime}\\
	&\text{if} ~ \boldsymbol\lambda_2 := -(1-w)\bK^{-1}(\rho)[\dfrac{\partial
	\varphi}{\partial \bu}]. \nonumber
\end{align}

For compliance minimization problems, since $\varphi = \bu^T\bff$ , the  $[\dfrac{\partial\varphi}{\partial \bu}] = \bff$ and $ \boldsymbol\lambda_2^T = (1-w)\bu^T(\rho)$. Thus,
\begin{align}
	\mathcal{L}'(\rho) &= w\MSSImax' (\rho) + (1-w)\bu^\mathrm{T}(\rho) \bK'(\rho)\bu(\rho) \nonumber\\&+ \lambda_1 \dfrac{1}{V_{\targ}}. \label{eq_Lag_primeCompliance}
\end{align}

To compute the maximum MSSI sensitivity field, we extend the surrogate model using a max pooling layer with a pool size as large as the domain size. This essentially allows the surrogate model to predict the peak MSSI value rather than the full MSSI field. Subsequently, we use automatic differentiation capabilities in $\mathsf{TensorFlow^{TM}}$ ~\cite{tensorflow2015whitepaper} to compute the change of the output of the NN (i.e., maximum MSSI) with respect to \textit{hypothetical} change in the input design variable (i.e., pseudo-density).  
Figure~\ref{fig_MSSIinChannel} illustrates the predicted MSSI field and the gradient field with respect to maximum MSSI value in the design domain, here the No-Go channel. 
Figure~\ref{fig_TOwithSurrogate} shows an overview of the proposed MSSI-aware TO framework.

%% file: Results.tex
\section{Results}\label{sec_results}

In this section, we demonstrate the effectiveness of our proposed method in generating MSSI-aware optimized designs, including experimental validation. The computations are on a desktop machine with
\textsf{Intel}$^{\small{\textregistered}}$
\textsf{Core}$^{\small{\texttrademark}}$i7-7820X CPU with 8 processors running
at 4.5 GHz, 32 GB of host memory, and an
\textsf{NVIDIA}$^{\small{\textregistered}}$
\textsf{GeForce}$^{\small{\textregistered}}$ GTX 1080 GPU with 2,560
\textsf{CUDA} cores and 8 GB of device memory. At every TO iteration, the design variables are updated using the Method of Moving Asymptotes (MMA)~\cite{svanberg1987method}, where the problem is approximated by a number of convex sub-problems which are solved using the Interior Point Method~\cite{nocedal2006numerical}.   

\begin{figure} [!h]
	\centering
	\includegraphics[width=\linewidth]{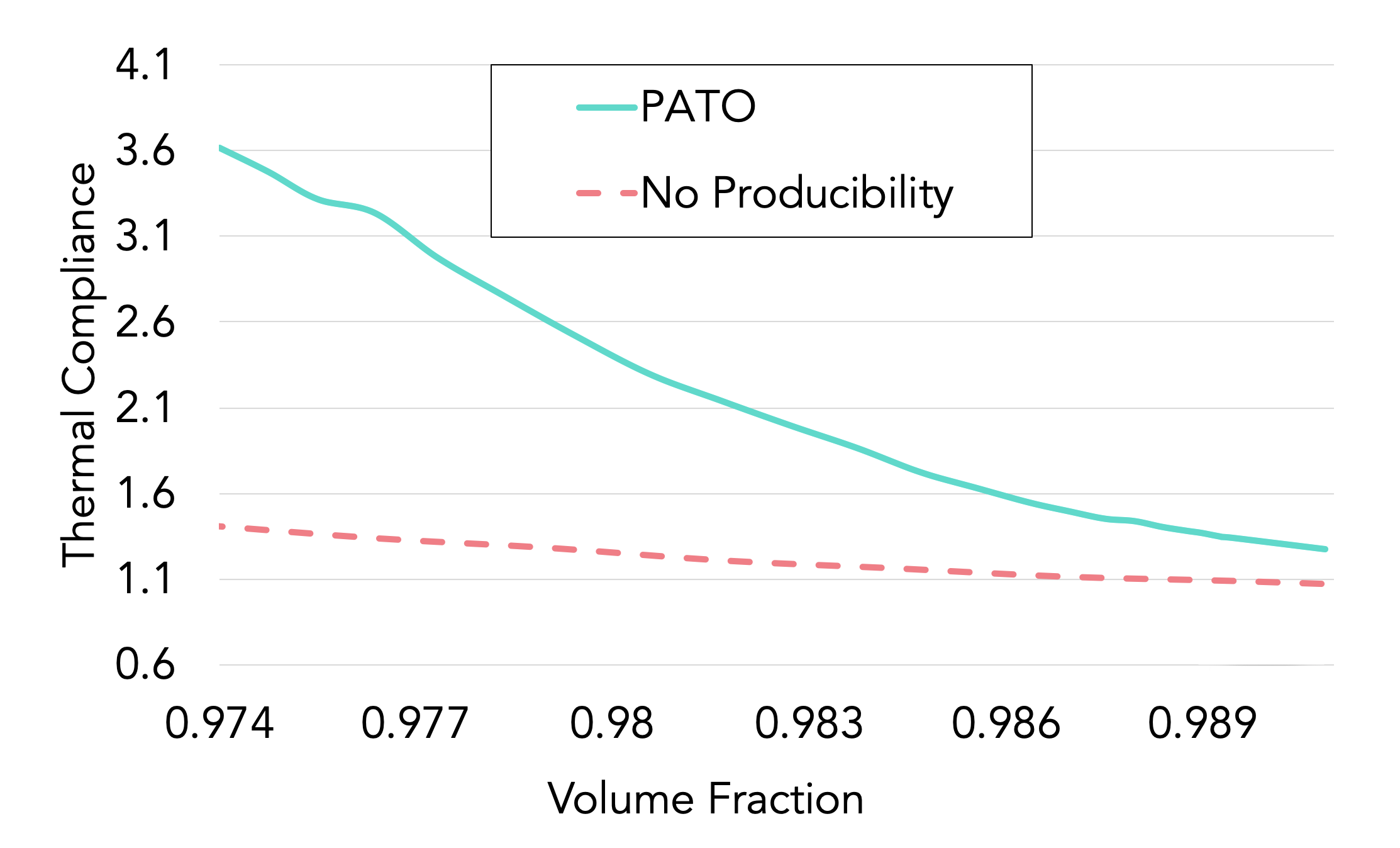}
	\caption{Thermal compliance at different volume fractions.}
	\label{fig_complianceGraphVer2}
\end{figure}

Figure~\ref{fig_MSSIConsVsUncVer1} illustrates the optimized designs at different volume fractions for TO without considering producibility ($w=0.0$) and the proposed PATO  ($w=0.95$) while minimizing thermal compliance under the loading condition of Fig.~\ref{fig_trainingData}a. The PATO designs are qualitatively different, where material is removed farther from the notch to reduce the maximum MSSI value.
\begin{figure} [!h]
	\centering
	\includegraphics[width=\linewidth]{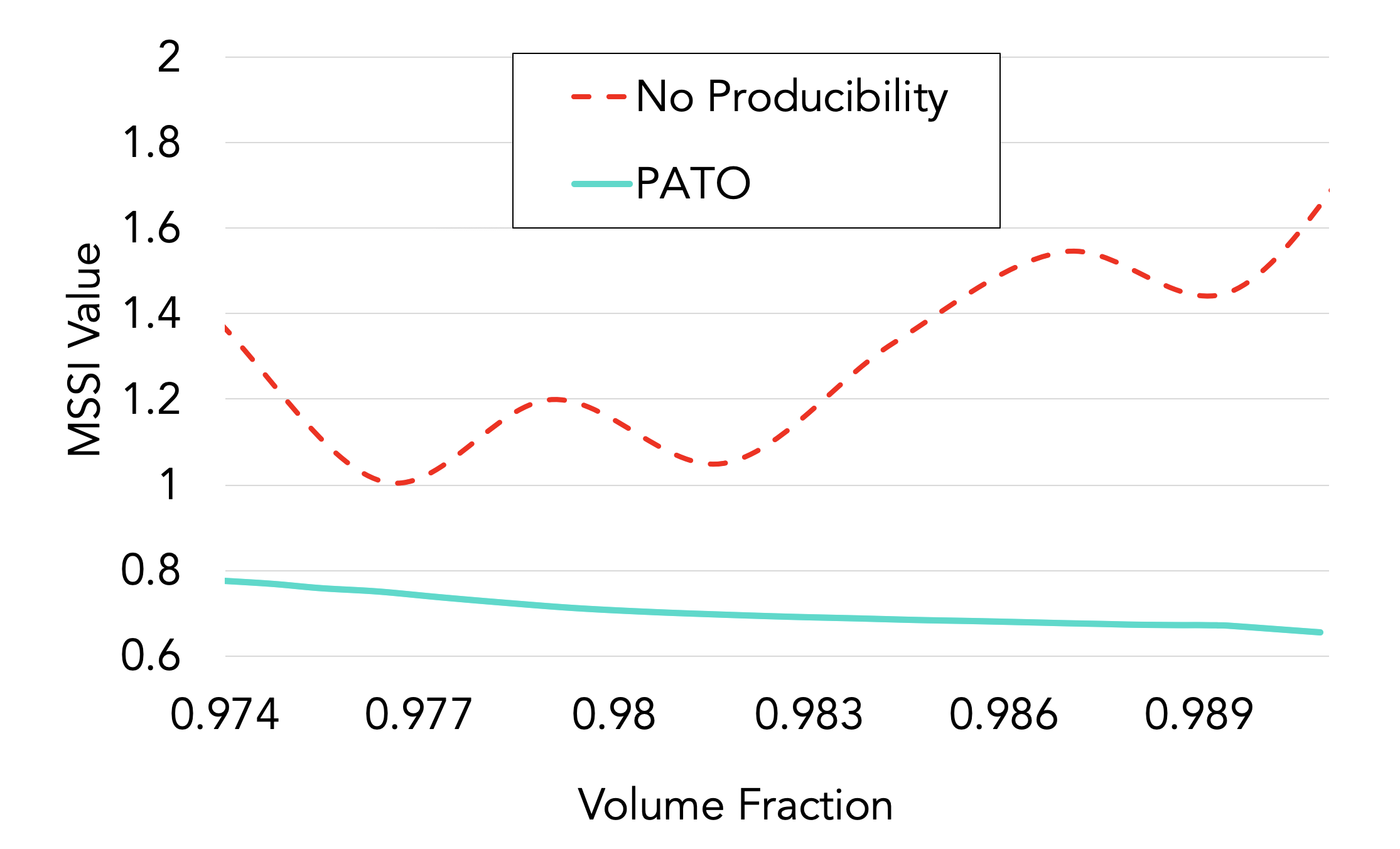}
	\caption{Maximum MSSI values at different volume fractions.}
	\label{fig_bndryMSSIGraphVer2}
\end{figure}
Figure~\ref{fig_bndryMSSIGraphVer2} shows the maximum MSSI values at different volume fractions. As expected, PATO designs consistently have lower maximum MSSI values. Figure~\ref{fig_complianceGraphVer2} illustrates the thermal compliance at different volume fractions. As expected, the  solutions found with PATO have higher thermal compliance values compared to those obtained without considering producibility. Thus, there is a trade-off between two  competing objectives, namely performance and producibility.

\begin{figure} [!h]
	\centering
	\includegraphics[width=0.95\linewidth]{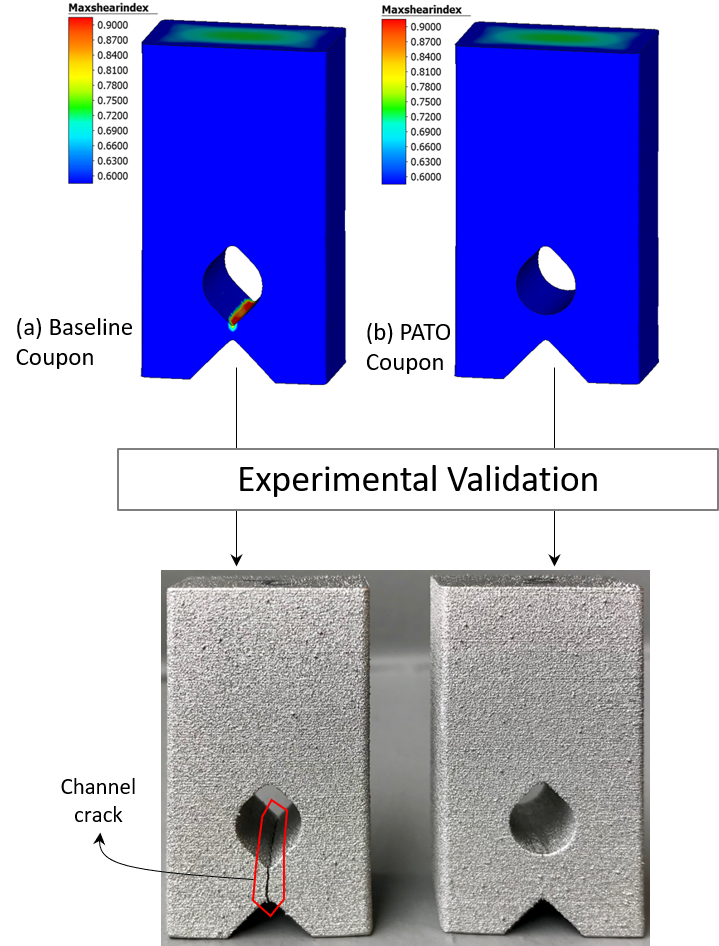}
	\caption{Comparing the MSSI values between the (a) Baseline No-Go coupon and (b) the optimal design discovered by PATO, along with pictures of experimental validation of the optimal design showing no cracks in corresponding printed coupons.}
	\label{fig_validation}
\end{figure}

Figure~\ref{fig_validation} shows a comparison in MSSI values between the baseline No-Go coupon and the PATO design discovered by application of our approach. It can be seen that the MSSI values are uniformly low across the channel as well as the overall coupon geometry for the optimized design, relative to values seen with the Baseline No-Go coupon. We also conducted experimental validation of our outcome by printing multiple coupons of the optimal design discovered by our approach. It is clear from the figure (bottom right), and based on a thorough inspection, that this coupon has no cracks either in the channel or any other region of the coupon, further validating that our approach converges to a design that is truly crack free.

%% file: Conclusion.tex
\section{Conclusions}
In this paper, we explored design manufacturability as it pertains to designs that crack upon additive manufacturing - a problem that is a challenge in the industry today for laser powder bed fusion of Ni-based superalloys used in hot gas path gas turbine components. We introduced a framework for producibility-aware topology optimization, called PATO, and demonstrated its ability to discover designs that are optimal from multiphysics perspective, while also ensuring their crack-free manufacturing. We showed the MSSI to be a reliable crack index, and how a deep convolutional neural network, with an Attention-based, 3D U-Net architecture, can be trained as a reliable surrogate of the time-complex build process simulation and accurately predict MSSI, given the geometry. Further, we showed how we employed automatic differentiation to directly compute the gradients of maximum MSSI with respect to the input design variables, using the surrogate, and augmented it with the performance-based sensitivity field in a standard topology optimization engine, to effectively optimize a design while considering the trade-off between weight, manufacturability, and multiphysics performance. We demonstrated the effectiveness of our PATO framework through benchmark studies in 3D as well as experimental validation. More specifically, the design suggested by PATO not only shows reduced values of MSSI, but this coupon when built did not have a crack, thereby validating the efficacy of our PATO framework to discover crack-free designs.

%% file: Acknowledgements.tex
\section*{Acknowledgments}
The information, data, or work presented herein was funded in part by the Advanced Research Projects Agency-Energy (ARPA-E), U.S. Department of Energy, under Award Number DE-AR0001203. The views and opinions of authors expressed herein do not necessarily state or reflect those of the United States Government or any agency thereof.